\documentclass[11pt,twoside]{article}

  \font\tenmib=cmmib10
\textfont"E=\tenmib \font\tenbsy=cmbsy10 \textfont"F=\tenbsy

\font\tbs=cmbsy10
\def\bnabla{\hbox{\tbs\char'162}}

\def\be{\begin{equation}}
\def\ee{\end{equation}}
\def\bea{\begin{eqnarray}}
\def\eea{\end{eqnarray}}

\usepackage{graphicx}

\usepackage{asp2006}

\begin{document}
\title{Five ideas on black hole accretion disks}
\author{Marek A. Abramowicz}
\affil{ Physics Department, G{\"o}teborg University, SE-412-96
G{\"o}teborg, Sweden\\ Copernicus Astronomical Center, Bartycka
18, 00-716 Warsaw, Poland }

\begin{abstract}
I review five of Bohdan Paczy{\'n}ski's ideas on black
hole accretion disk theory. They formed my understanding of the
subject and often guided intuition in my research. They are
fundamentally profound, rich in physical consequences,
mathematically elegant and clever, and in addition are useful in
several technically difficult practical applications.
\end{abstract}


\section{Introduction} \label{section-introduction}


\subsection{Motivation: the five ``easy'' pieces}

This review article does not give a full (or even coherent)
description of black hole accretion disk theory. Instead, it
only reflects briefly on a few of Bohdan Paczy{\'n}ski's particularly
important contributions to the subject. Bohdan had several
brilliant ideas about black hole accretion.  Today, most of them are
mainstream and firmly accepted. One is
still considered controversial.

{\it 1. The Potential:~} For free particles, both Newton's and
Einstein's orbital dynamics are described by the same principle,
and indeed the same equation: the orbital frequency follows from
the first derivative, and the epicyclic frequencies follow from
the second derivatives of the ``effective potential''. Therefore,
as Bohdan pointed out, by a proper definition of an artificial
Newtonian potential, one should be able to accurately describe
(formally in Newton's theory!) the fully general relativistic
orbital motion. He then guessed the form of the
``Paczy{\'n}ski-Wiita'' potential which immediately became a great
success. The potential is simple and very practical in numerous
applications.

{\it 2. The Doughnut:~} One is often interested in phenomena that
occur on a ``dynamical'' timescale $t_{*}$ much shorter than the
``viscous'' time $t_{\cal L}$ needed for angular momentum
redistribution, and the ``thermal'' time $t_{\cal S}$ needed for
entropy redistribution. In modeling such cases, Bohdan noticed,
the angular momentum and entropy distributions may be considered
{\it free functions}. Therefore, when studying processes
with $t_* \ll t_0 = {\rm Min}(t_{\cal L}, t_{\cal S})$, one may
ad hoc {\it assume} angular momentum and entropy distributions. A
constant angular momentum is the simplest possible assumption. It
was used by Bohdan and his Warsaw team to construct the ``Polish
Doughnuts'', i.e. super-Eddington thick accretion disks.

{\it 3. The Funnel:~} In Warsaw we found that a long,
narrow empty funnel forms along the axis of rotation of a Polish
doughnut. It collimates radiation to hyper-Eddington fluxes.

{\it 4. The Roche Lobe:~} In equilibrium, the isobaric surfaces of
the accretion disk coincide with the surfaces of constant
effective potential, called the ``equipotential'' surfaces. Near a
black hole one of the equipotential surfaces, called the ``Roche
lobe'', self-crosses along the ``cusp'', i.e. a circle
$r=r_{cusp}$ located between the innermost stable circular orbit
at $r = r_{ISCO}$, and the marginally bound orbit at $r=r_{mb}$.
Bohdan's knowledge of close binaries helped him to realize that
the black hole Roche lobe overflow must induce dynamical mass loss
from the disk. Thus, for $r < r_{cusp}$ accretion is not caused by
``stresses'' that remove angular momentum, but by the strong
gravity. The ``inner edge'' of an accretion disk is at the ISCO
only for accretion flows with very small ${\dot M}/{\dot M}_{Edd}$
that are radiatively efficient (Shakura-Sunyaev). He understood
that the efficiency of accretion $\eta$ (defined by $L = \eta
{\dot M}c^2$) equals the Keplerian binding energy at the location
of the cusp: when $r_{cusp} = r_{ISCO}$, efficiency takes its
maximal possible value, but it goes to zero when $r_{cusp}$
approaches $r_{mb}$. All radiatively inefficient accretion flows
(RIAFs) like advection-dominated accretion flows (adafs) and slim
disks have their inner edges close to $r_{mb}$.

{\it 5. The Inner Torque:~} Against the view of an influential part
of the black hole community, Bohdan argued that when the accretion
disk is vertically very thin, for very small accretion rates the
viscous or magnetohydrodynamic (MHD) torque at the inner edge
should be vanishingly small.

\begin{figure}
  \includegraphics[width=.95\textwidth]
  {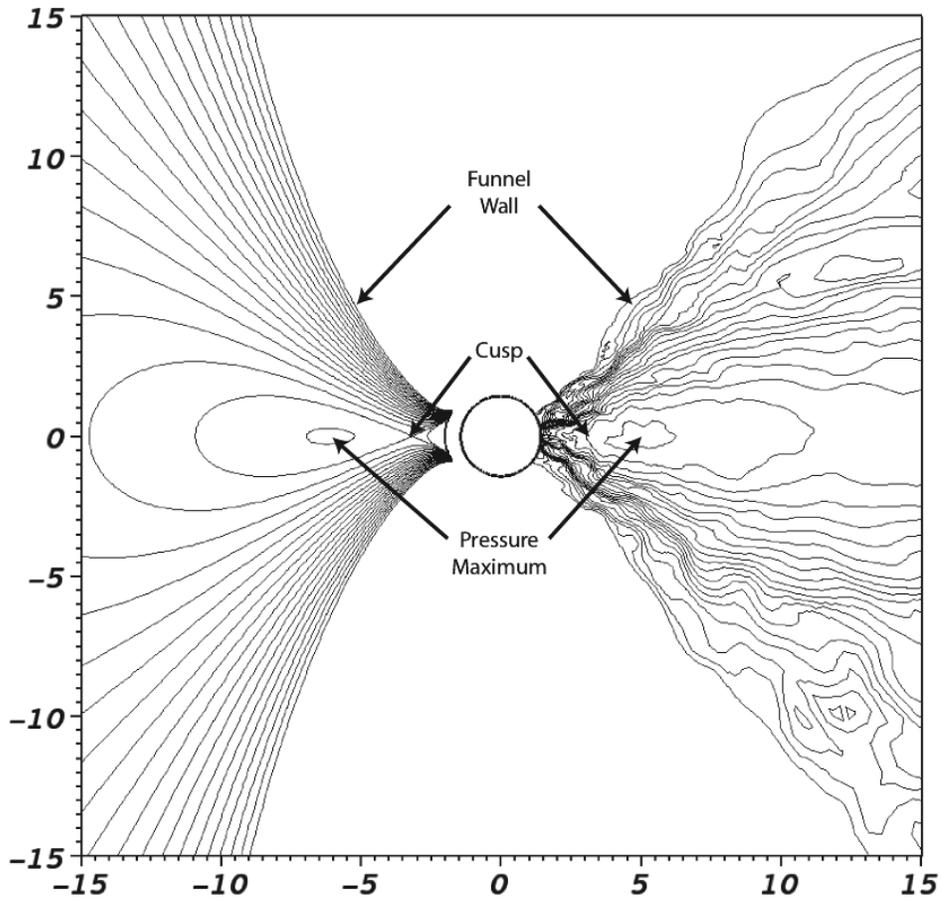}
\caption{Isobaric surfaces in a black hole accretion flow.
      Left: according to a very simple, analytic
      Polish doughnut model. Right: according to a
      state-of-art, full 3D MHD numerical simulation.
      Figure taken from the {\it Living Review on black hole accretion
      disks} by Abramowicz \& Fragile, in preparation.}
\label{fragile}
\end{figure}

\subsection{Analytic models and computer simulations}

Bohdan was a master of massive supercomputing methods in
astrophysics. Some of the computer codes that he developed, e.g.
for stellar structure and evolution, are the standard tools in the
field. However, although supercomputer MHD simulations are
a large part of black hole accretion research today, Bohdan
was not himself involved in this activity. Indeed, all Bohdan's
ideas that I discuss here are based on simple analytic models. One
may ask --- why? This I do not know, but I can think of a
possible reason. The present-day MHD simulations do not address
the fundamental problems that the theory of black hole accretion
faces. Our understanding of black hole accretion rests on
analytic models. Indeed, all the fundamental features of black
hole accretion flows that have been calculated in MHD
simulations, were {\it previously well understood} in terms of
analytic models. It seems that today, and in the foreseeable future,
the super-computer simulations will keep confirming rather than
solving or discovering. Today, the most important message of the
supercomputer simulations seems to be that the approximations and
simplifications adopted in the analytic models have been rather
wisely chosen (see Figure \ref{fragile} to illustrate this
point).

\subsection{Notation and units}

In this review, I use the Kerr metric in the standard spherical
Boyer-Lindquist coordinates $t, \phi, r, \theta$ with the $+---$
signature. The Kerr metric is stationary and axially symmetric,
\begin{equation}
\label{the-metric} ds^2 = g_{tt}\,dt^2 + 2g_{t\phi}\,dt\,d{\phi} +
g_{\phi\phi}\,d{\phi}^2 + g_{rr}\,dr^2 +
g_{\theta\theta}\,d{\theta}^2,
\end{equation}
i.e. its metric tensor components are (known) functions of radial
and polar coordinates  $r$, $\theta$, the mass $M$ and the
dimensionless spin parameter $\vert a \vert < 1$,
\begin{equation}
\label{the-kerr-metric} g_{ik} = g_{ik}(r, \theta; M, a).
\end{equation}
However, in all specific examples, I will use a much simpler
Schwarzschild metric that describes spacetime geometry of a
non-rotating ($a=0$) black hole,
\begin{equation}
\label{the-schwarzschild-metric} ds^2 = \left(1 -
\frac{r_G}{r}\right)dt^2 - \left(1 - \frac{r_G}{r}\right)^{-1}dr^2
- r^2\left[ d\theta^2 + \sin^2\theta\,d{\phi}^2\right],
\end{equation}
with the gravitational radius $r_G$ is defined by,
\begin{equation}
\label{gravitational-radius} r_G \equiv \frac{2GM}{c^2}
\end{equation}
My apologies to Bohdan; here I often but not always use the $c = 1 = G$
units he greatly disliked.


\section{The Potential} \label{section-potential}


According to Newton's theory of gravity, the motion of a free
particle in a stationary, axially symmetric gravitational
potential $\Phi(r, \theta)$, is confined to a plane, which may be
chosen as the ``equatorial plane'', $\theta = \pi/2$. Circular
orbits in the equatorial plane are characterized by
\begin{equation}
\label{circular-orbits} r = r_0 = {\rm const}, ~~~ \theta =
\theta_0 = \frac{\pi}{2} = {\rm const},
\end{equation}
and the slightly non-circular, slightly off-plane orbits are
characterized by
\begin{equation}
\label{circular-orbits-deviations} \delta r (t) \equiv r(t) - r_0
\ll r_0, ~~~ \delta \theta (t) \equiv \theta(t) - \theta_0 \ll
\theta_0.
\end{equation}
The specific (per unit mass) energy ${\cal E} = [V^2 +
(v_{\phi})^2]/2 + \Phi$ and the specific angular momentum ${\cal
L} = r\,v_{\phi}$ are constants of motion. Here $V^2 = (v_r)^2 +
(v_{\theta})^2 \equiv (\delta {\dot r})^2 + (\delta {\dot
\theta})^2 \ll (v_{\phi})^2$. One defines the effective potential
\begin{equation}
\label{newton-effective-potential}
 {\cal U}(r, \theta, {\cal
L}) = \Phi(r, \theta) + \frac{{\cal L}^2}{2\,r^2\sin^2\theta},
\end{equation}
and writes,
\begin{equation}
\label{effective-potential} \frac{1}{2}V^2 = {\cal E} - {\cal
U}(r, \theta, {\cal L}).
\end{equation}
The Keplerian angular momentum distribution ${\cal L} = {\cal
L}_K(r)$ in strictly circular orbits (\ref{circular-orbits})
follows from the condition,
\begin{equation}
\label{effective-potential-first} \left( \frac{\partial {\cal
U}}{\partial r}\right )_{\cal L} = 0,
\end{equation}
while the deviations (\ref{circular-orbits-deviations}) obey the
simple harmonic oscillator equation\footnote{When deriving the
``radial'' equation, one puts $V = dr/dt = \delta {\dot r}$ in
(\ref{effective-potential}), and when deriving the ``vertical''
equation one puts $V = d\theta/dt = \delta {\dot \theta}$ in
(\ref{effective-potential}).}
\begin{equation}
\label{simple-harmonic} \delta {\ddot r} + \omega^2_r\, \delta r =
0, ~~~ \delta {\ddot \theta} + \omega^2_{\theta}\,\delta \theta =
0,
\end{equation}
with the ``radial'' $\omega_r$, and the ``vertical''
$\omega_{\theta}$ epicyclic frequencies given by,
\begin{equation}
\label{effective-potential-second} \omega_r^2 = \left(
\frac{\partial^2 {\cal U}}{\partial r^2}\right )_{\cal L},
~~\omega_{\theta}^2 = \left( \frac{\partial^2 {\cal U}}{\partial
\theta^2}\right )_{\cal L}.
\end{equation}
So much for the Newtonian case. The fully relativistic case is
remarkably similar. The relativistic analog of the Newtonian
condition for strictly circular orbits (\ref{circular-orbits}) may
be written as a condition for the components of a particle's
four-velocity $u^i = dx^i/ds$,
\begin{equation}
\label{four-velocity-circular} u^i = (u^t, u^{\phi}),~~~ u^r = 0,
~~~ u^{\theta} = 0.
\end{equation}
The nearly-circular orbits are characterized by,
\begin{equation}
\label{four-velocity} u^i = (u^t, u^{\phi}, u^r, u^{\theta}),
~~~(u^r, u^{\theta}) \ll (u^t, u^{\phi}).
\end{equation}
As in Newton's theory, the energy and angular momentum are two
constants of the free (i.e.\ geodesic) motion. Their standard
definitions read ${\cal E}^* = u_t$ ${\cal L}^* = -u_{\phi}$, but
for our purpose it is convenient to rescale them according to,
\begin{equation}
\label{redefinition-constant-motion} {\cal E}_{\rm Kerr} =
\ln{\cal E}^*, ~~~{\cal L}_{\rm Kerr} = \frac{{\cal L}^*}{{\cal
E}^*}.
\end{equation}
Obviously,  ${\cal E}_{\rm Kerr}$ and ${\cal L}_{\rm Kerr}$ are
also constants of geodesic motion. It is also convenient to define
the small ``deviation'' velocity $V$ by its square,

\begin{equation}
\label{definition-velocity} V^2_{\rm Kerr} = 2\,\ln \left[ 1
-g_{rr}(u^r)^2 - g_{\theta\theta}(u^{\theta})^2\right] \ll 1,
\end{equation}
which is obviously always positive because $-g_{rr} > 0$ and
$-g_{\theta\theta}
> 0$. Then, most importantly, one introduces the effective potential,
\begin{equation}
\label{definition-kerr-potential} {\cal U}_{\rm Kerr} =
-\frac{1}{2} \ln \left( g^{tt} - 2{\cal L}g^{t\phi}
 + {\cal L}^2\,g^{\phi \phi}\right).
\end{equation}
With these definitions, one may easily cast the general equation
$u_i u_k g^{ik} = 1$ into the form
\begin{equation}
\label{kerr-motion} \frac{1}{2}V^2_{\rm Kerr} = {\cal E}_{\rm
Kerr} - {\cal U}_{\rm Kerr}(r, \theta, {\cal L}_{\rm Kerr}),
\end{equation}
which is identical with the corresponding Newtonian equation
(\ref{effective-potential}). All quantities that appear in
(\ref{kerr-motion}) have the same physical meaning as the
corresponding Newtonian quantities, and go to the right limits
when $v_{\phi}/c \ll 1$, $V/c \ll 1$, $(2GM/c^2)/r = r_G/r \ll 1$.
\begin{figure}
\begin{center}
\label{figure-paczynski-wiita-quotations}
  \includegraphics[width=7cm]
  {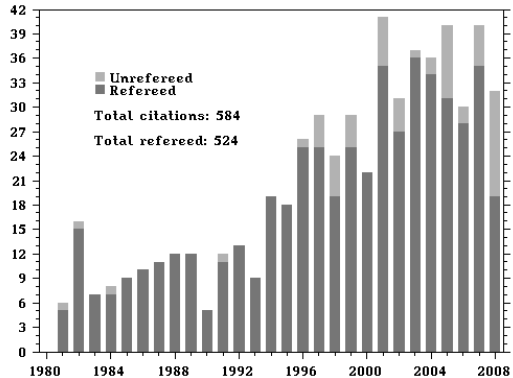}
\end{center}
\caption{The Paczy{\'n}ski-Wiita potential
(\ref{paczynski-wiita-gravity}) was first introduced in the paper
by \citet*{pac-wii-1980}. This is the most frequently cited paper by
Bohdan on the subject of the black hole accretion. It ranks in the
fourth place on the list of his most frequently cited papers. The
figure and citation numbers were taken from
the SAO/NASA ADS Astronomy Query service on June 6, 2008. }
\end{figure}
{\it The same equations have the same solutions}\footnote{This
well-known remark by Richard Feynman is ``unsourced'' according to
wikiquote.} and therefore one may just use the same procedure as
in Newton's theory to get the Keplerian angular momentum from
the first derivative of the effective potential, i.e. from
equation (\ref{effective-potential-first}), and the epicyclic
frequencies from the second derivatives of the potential, i.e.
from equation (\ref{effective-potential-second}). This is indeed
the standard, text-book way used in general relativity. For
example, in the special case of a non-rotating (i.e.
Schwarzschild) black hole, the relativistic effective potential
(\ref{definition-kerr-potential}) takes the form (on the
equatorial plane),
\begin{equation}
\label{potential-schwarzschild} {\cal U}_{\rm Sch} = - \frac{1}{2}
\ln \left[ \left( 1 - \frac{r_G}{r} \right)^{-1} - \frac{{\cal
L}^2}{r^2} \right],
\end{equation}
and demanding that its first derivative is zero leads to the
expression for the Keplerian angular momentum,
\begin{equation}
\label{kepler-schwarzschild} {\cal L}_K^2 = \frac{GM\,r^3}{(r -
r_G)^2}.
\end{equation}
Bohdan noticed that the same expression for the Keplerian angular
momentum one gets for the effective potential given by the formula
(on the equatorial plane),
\begin{equation}
\label{paczynski-wiita} {\cal U}_{PW} = - \frac{GM}{r - r_G} +
\frac{{\cal L}^2}{2r^2}.
\end{equation}
The second term on the right hand side, ${\cal L}^2/2r^2$, is the
standard Newtonian centrifugal part of the effective potential.
The first term should be therefore identified with a gravitational
potential,
\begin{equation}
\label{paczynski-wiita-gravity} \Phi_{PW} = - \frac{GM}{r - r_G},
\end{equation}
which is indeed the celebrated ``Paczynski-Wiita'' or
``pseudo-Newtonian'' potential, first introduced by Bohdan in his
paper with Paul Wiita \citep*{pac-wii-1980}.


\section{The Doughnut} \label{section-doughnut}


Several interesting phenomena that occur in black hole
accretion disks have characteristic timescales $t_*$ much
shorter than the time $t_{\cal L}$ needed to redistribute the
angular momentum, or the time $t_{\cal S}$ needed to redistribute
the entropy
\begin{equation}
\label{short-timescales}
t_{*} \ll t_0 = {\rm Min}\left(t_{\cal
L}, t_{\cal S}\right)
\end{equation}
During the time $t_*$, distributions of angular momentum and
entropy,
\begin{equation}
\label{distribution-momentum-entropy} {\cal L} = {\cal L}(r,
\theta), ~~~{\cal S} = {\cal S}(r, \theta)
\end{equation}
do not change. The functions (\ref{distribution-momentum-entropy})
cannot be unambiguously calculated from first principles, as
they depend on unknown initial conditions, and on unknown
details of dissipative processes such as e.g. turbulence,
convection, or radiative transfer. Therefore, when they {\it are}
calculated, for example with the help of lengthy and difficult
numerical simulations, results {\it are arbitrary}, at least to
some degree.
\begin{equation}
\label{results-arbitrary} \left( \begin{array}{c}{\it step ~1}\\
~\\{\rm arbitrary}\\
{\rm assumptions}\end{array}\right)
\rightarrow \left( \begin{array}{c}{\it step ~2}\\~\\{\rm difficult}\\
{\rm calculations}\end{array}\right) \rightarrow \left(
\begin{array}{c}{\it step ~3}\\~\\{\cal L} = {\cal L}(r, \theta)\\{\cal
S} = {\cal S}(r, \theta)\end{array}\right)
\end{equation}
Paczy{\'n}ski argued that there is no obvious reason to believe
that it would be easier to guess what needs to be assumed in
step 1 than to assume (guess) the result in the step 3. In the
late 1970s and early 1980s he and his collaborators explored
accretion disk models in which distributions of angular momentum
and entropy  (\ref{distribution-momentum-entropy}) were assumed ad
hoc.

For a perfect-fluid matter, $T^i_{~k} = (p + \epsilon)u^i\,u_k -
p\delta^i_{~k}$, with $p$ and $\epsilon$ being the pressure and
the total energy density. Equations of mass conservation and
energy-momentum conservation take the form,
\begin{equation}
\label{mass-energy-conservation} \nabla_i (u^i \rho) = 0,
~~~\nabla_i\,T^i_{~k} = 0,
\end{equation}
where $\rho$ is the rest-mass density. From these equations one
deduces that two Bernoulli-type quantities are constant along the
stream-lines of the perfect fluid matter moving in a stationary
and axially symmetric spacetime
(\ref{the-metric})-(\ref{the-kerr-metric}).
\begin{equation}
\label{two-Bernoulli} {\cal B}^* = \left( \frac{p +
\epsilon}{\rho}\right)\,u_t,~~~{\cal L}^* = - \left(\frac{p +
\epsilon}{\rho}\right)\,u_{\phi}
\end{equation}
Thus, the purely kinematic quantity, defined previously as the
specific angular momentum for particles,
\begin{equation}
\label{Bernoulli-specilic-angular} {\cal L} = \frac{{\cal
L}^*}{{\cal B}^*} = - \frac{u_{\phi}}{u_t},
\end{equation}
is also constant along fluid's stream-lines. If the fluid moves
along circular stream-lines (\ref{four-velocity-circular}), the
equation of motion $\nabla_i\,T^i_{~k} = 0$ yields (c.f. equation
(\ref{definition-kerr-potential})),
\begin{equation}
\label{Euler} \frac{\nabla_i p}{p + \epsilon} = -\frac{1}{2}
\frac{\nabla_i\,g^{tt} - 2{\cal L}\,\nabla_i g^{t\phi} + {\cal
L}^2\,\nabla_i g^{\phi\phi}}{g^{tt} - 2{\cal L}\,g^{t\phi} + {\cal
L}^2\,g^{\phi\phi}} = \left( \frac{\partial {\cal U}_{\rm
Kerr}}{\partial x^i}\right )_{\cal L},
\end{equation}
which may be transformed into,
\begin{equation}
\label{von-Zeipel} \frac{\nabla_i p}{p + \epsilon} = \nabla_i
{\cal U} + \frac{\Omega\,\nabla_i {\cal L}}{1 - {\cal L}\,\Omega}.
\end{equation}
Here $\Omega = u^{\phi}/u^t$ is the angular velocity, related to
the angular momentum ${\cal L}$ by the following formulae, which are
very useful in practical calculations.
\begin{equation}
\label{angular-velocity-momentum} {\cal L} = -
\frac{\Omega\,g_{\phi\phi} + g_{t\phi}}{\Omega\,g_{t\phi} +
g_{tt}}, ~~ \Omega = - \frac{{\cal L}\,g_{tt} + g_{t\phi}}{{\cal
L}\,g_{t\phi} + g_{\phi\phi}}
\end{equation}

\subsection{The von Zeipel fluid tori}

Suppose that $p = p(\epsilon)$, as it would be for isentropic,
polytropic or barytropic fluid. It this case, the left hand side
of equation (\ref{von-Zeipel}) is a perfect gradient, and so must
be the right hand side, and this is possible when ${\cal L} =
{\cal L}(\Omega)$. Similarly, if ${\cal L} = {\cal L}(\Omega)$,
then the right hand side is a gradient, and so must be the left
hand side. This statement is known as the von Zeipel condition,
\begin{equation}
\label{von-zeipel-condition} \left[
\begin{array}{c} p = p(\epsilon) \end{array}\right]
\Leftrightarrow \left[
\begin{array}{c}{\cal L} = {\cal L}(\Omega)
\end{array}\right],
\end{equation}
or, the $p(r,\theta) =$ const surfaces conincide with those of
$\epsilon(r,\theta) =$ const, {\it if and only if} the surfaces
${\cal L}(r,\theta) =$ const coincide with those $\Omega(r,\theta)
=$ const\footnote{The best known Newtonian version of the von
Zeipel condition states that for a barytropic fluid $p =
p(\epsilon)$, both angular velocity and angular momentum are
constant on cylinders, $\Omega = \Omega(R)$, ${\cal L} = {\cal
L}(R)$, with $R = r\sin\theta$ being the distance from the
rotation axis.}.

The simplest example of a fluid fulfilling the von Zeipel
condition (\ref{von-zeipel-condition}) is
\begin{equation}
\label{constant-momentum-entropy}  {\cal L}(r, \theta) = {\cal
L}_0 = {\rm const}, ~~~{\cal S}(r, \theta) = {\cal S}_0 = {\rm
const}.
\end{equation}
In this case, equation (\ref{von-Zeipel}) has a trivial first
integral
\begin{equation}
\label{first-integral} W(p) = {\cal U}_{\rm Kerr}(r, \theta),
\end{equation}
where $W(p)$ is the enthalpy, defined for isentropic fluids
(\ref{constant-momentum-entropy}) by
\begin{equation}
\label{enthalpy}  W = W(p) = \int \frac{dp}{p + \epsilon}.
\end{equation}
Note that (\ref{first-integral}) gives the location of the
isobaric surfaces in terms of an {\it explicit} analytic
function of coordinates $r, \theta$. From (\ref{Euler}) it follows
that the pressure maximum is located at the circle $r = r_0$,
$\theta = \theta_0 = \pi/2$, given by the condition,
\begin{equation}
\label{pressure-maximum} \left( \frac{\partial {\cal U}_{\rm
Kerr}}{\partial r}\right )_{\cal L} = 0 = \left( \frac{\partial
{\cal U}_{\rm Kerr}}{\partial \theta}\right )_{\cal L}.
\end{equation}
Let us express the effective potential near the pressure maximum
circle $r_0, \theta_0$ by
\begin{equation}
\label{pressure-maximum-potential} \Delta {\cal U} = {\cal U}_{\rm
Kerr}(r, \theta) - {\cal U}_{\rm Kerr} (r_0, \theta_0).
\end{equation}
Because the first derivative of the effective potential vanishes
at the maximum pressure circle, and the second derivatives equal
to the epicyclic frequencies (c.f. equation
(\ref{effective-potential-second})), one concludes that the shape
of a ``slender'' torus, i.e. with $\Delta r \equiv r - r_0 \ll
r_0$, $\Delta \theta \equiv \theta - \theta_0 \ll \theta_0$, is
given by $\Delta {\cal U}=~$const, with
\begin{figure}[ht*]
\begin{center}
\includegraphics[width=.5\textwidth]{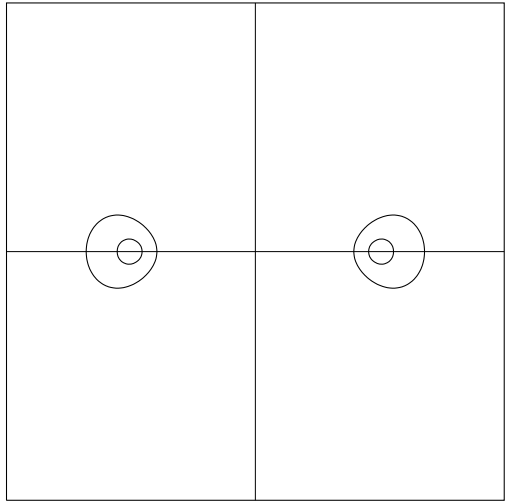}
\end{center}
\caption[]{Contours of equipotential surfaces on the meridional
cross section of a torus with constant angular momentum, exactly
calculated from (\ref{first-integral}), according to Newton's
theory with the $-GM/r$ gravitational potential. The contours
approach circles around the locus of the maximum pressure, at the
$r=r_0$, $\theta=\theta_0=\pi/2$ circle, which agrees with the
approximate ``slender torus'' solution given by equation
(\ref{slender-torus-shape}).}
\end{figure}
\begin{equation}
\label{slender-torus-shape} \Delta {\cal U}(r, \theta) =
\frac{1}{2}\left[ \omega_r^2\,(\Delta r)^2 +
\omega_{\theta}^2\,(\Delta \theta)^2 \right] + {\cal
O}^3(\varepsilon),~~~\varepsilon =\frac{\Delta r}{r_0} \ll 1.
\end{equation}
Equation (\ref{slender-torus-shape}) is valid both in Newton's and
Einstein's theory. In Newton's theory with $-GM/r$ gravitational
potential, one has $\omega_r = \omega_{\theta}$. Thus, in this
case, one concludes from equation (\ref{slender-torus-shape}) that
slender constant-angular-momentum fluid tori have {\it circular}
cross-sections. This additional symmetry was first noticed by
Bohdan in his paper with Jurek Madej \citep*{mad-pac-1977}, and
later used by \cite{bla-1985} to calculate normal modes of oscillations
of slender tori. Blaes demonstrated that there exists a set of
non-axisymmetric (i.e. with $m \not = 0$) global modes $\delta
\Psi_{mn} \sim \exp (-i \sigma_{mn} t + m \phi)$ with the
eigenfrequencies,
\begin{equation}
\label{omer-unstable} \sigma_{mn} = - m \Omega_K \left [ 1 + i
\epsilon \left ( { 3 \over {2n + 2}} \right )^{1\over 2} \right] +
{\cal O}^2 (\epsilon), ~~n=1,2,3,...
\end{equation}
\noindent Because ${\Im}(\sigma_{mn}) < 0$ these oscillations are
unstable. The growth rate of the instability is $\sim m\Omega_K$,
and thus the instability is a {\it dynamical} one. Indeed, this is
the famous Papaloizou-Pringle instability, discovered in the
seminal paper by John Papaloizou and James Pringle
\citep*{pap-pri-1984}.

\begin{figure}[ht*]
\begin{center}
\includegraphics[width=.325\textwidth]{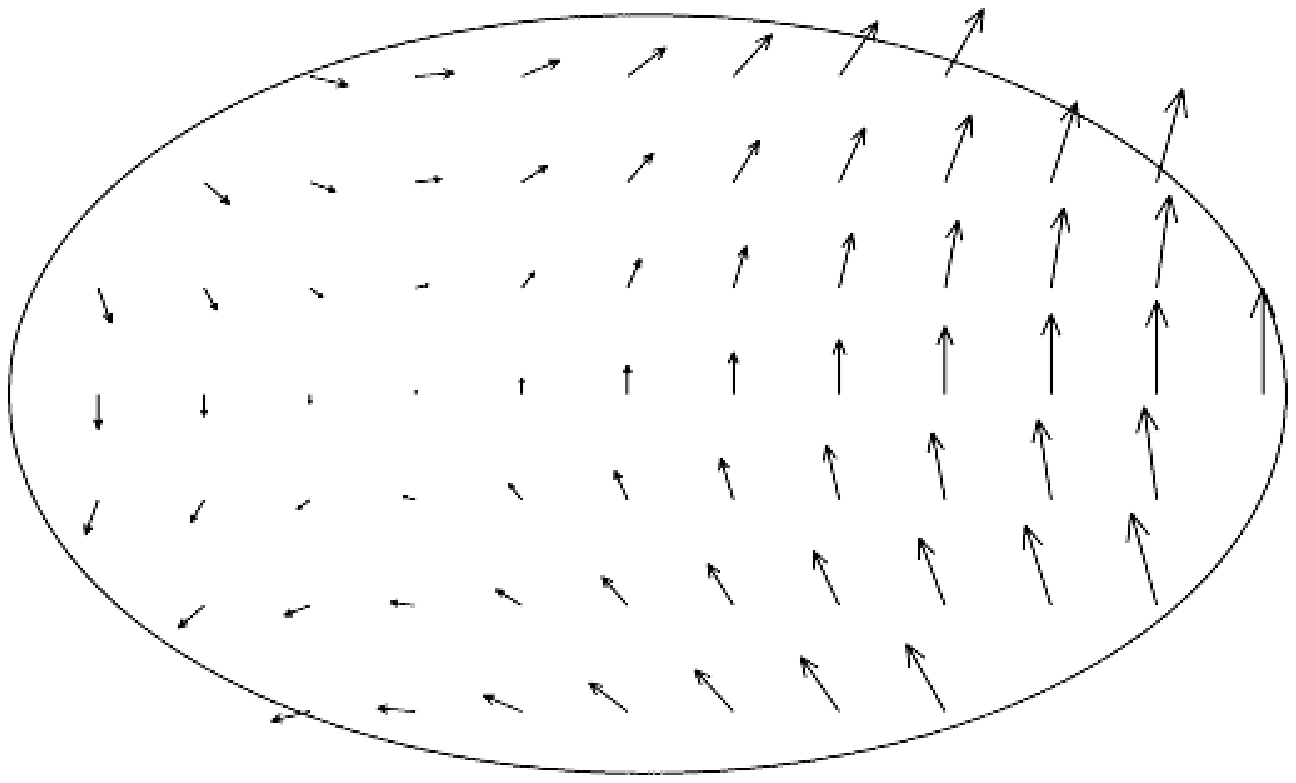}
\hfill
\includegraphics[width=.325\textwidth]{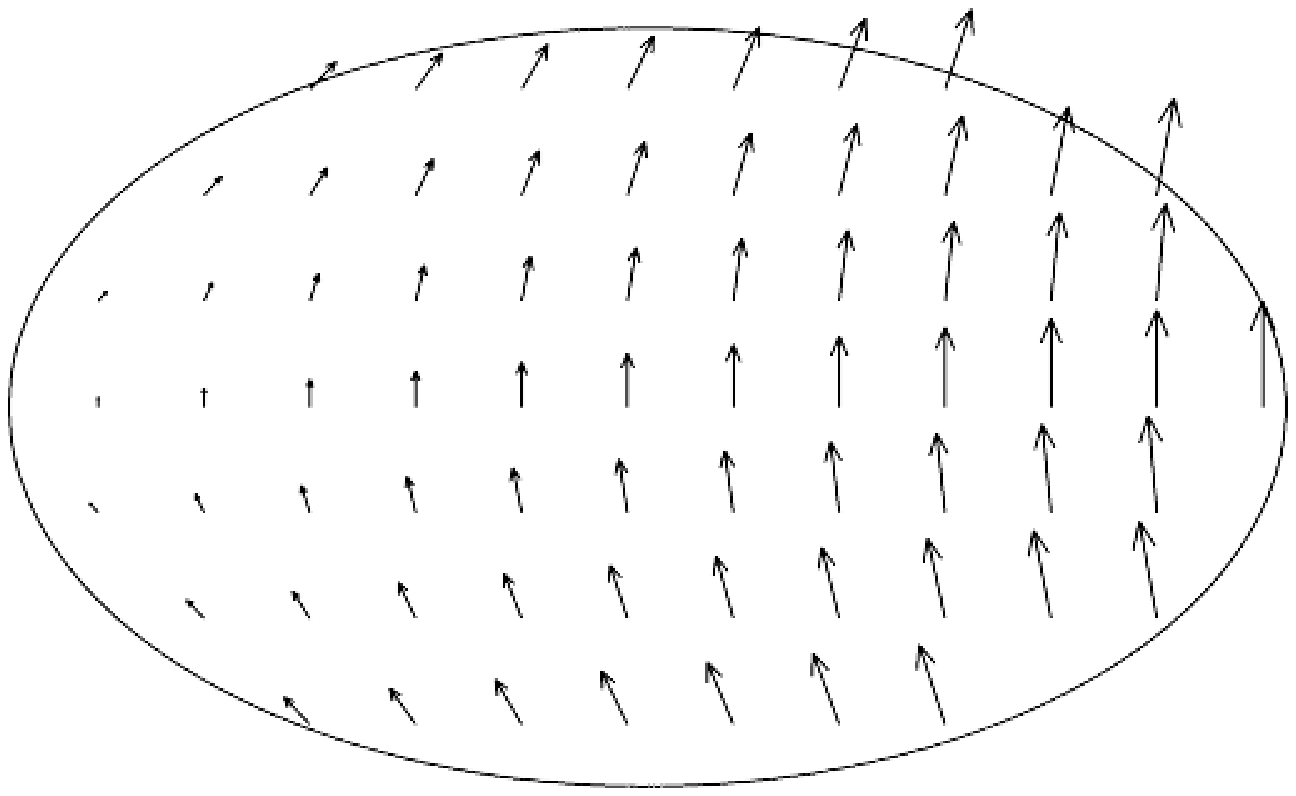}
\hfill
\includegraphics[width=.325\textwidth]{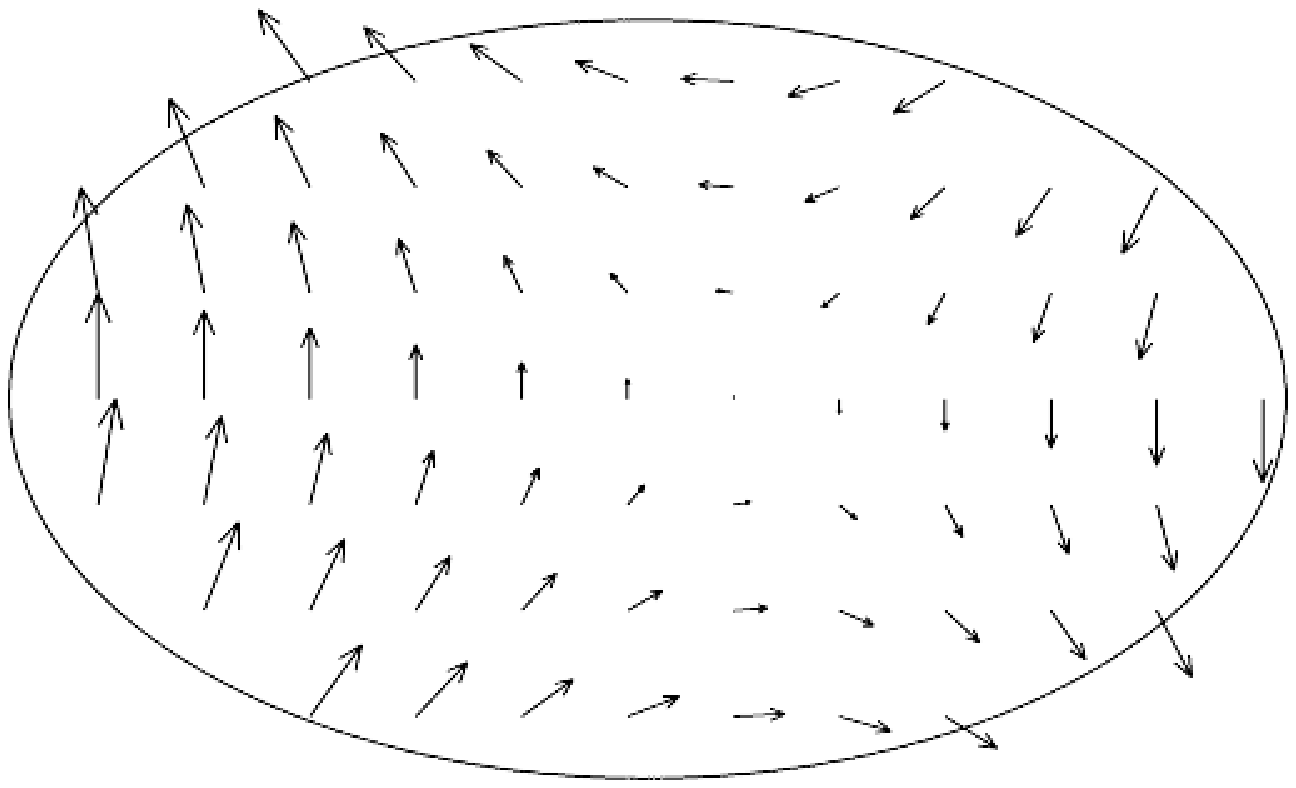}
\end{center}
\caption[]{Velocity pattern for the vertical epicyclic mode of
oscillation of a slender torus around a Kerr black hole as
calculated from an analytic model by \cite{str-sra-2008}.}
\label{odele-vertical}
\end{figure}

In the last twenty years, studies of slender tori oscillations have
brought a lot of important results. Today, these results are concentrated in
non-linear excitation, damping, and resonances of global epicyclic
modes that are thought to be relevant to the 3:2 twin peak QPOs.
The velocity pattern for the vertical epicyclic mode of a slender
torus in the Kerr geometry is shown in Figure \ref{odele-vertical},
taken from \cite{str-sra-2008}.

\subsection{General case}

\Citet{jar-1980} discuss a general, non-barytropic case of angular
momentum and entropy distributions (\ref{distribution-momentum-entropy}).
They start by noting that if $\theta = \theta(r)$ is the equation
for the isobaric surface $p(r, \theta) =$ const, then obviously
\begin{equation}
\label{equipressure-shape-first} \frac{d\theta}{dr} =
-\frac{{\partial_r}p}{\partial_{\theta}p}.
\end{equation}
They further note that the right hand side of the above
equation may be calculated from (\ref{Euler}), if one writes it
twice for $i = r$ and $i = \theta$, divides the two
equations, and changes the sign,
\begin{equation}
\label{master} - \frac{\partial_r\,p}{\partial_{\theta}\,p} = -
\frac{\partial_r\,g^{tt} - 2{\cal L}\,\partial_r g^{t\phi} + {\cal
L}^2\,\partial_r g^{\phi\phi}}{\partial_{\theta}\,g^{tt} - 2{\cal
L}\,\partial_{\theta} g^{t\phi} + {\cal L}^2\,\partial_{\theta}
g^{\phi\phi}}.
\end{equation}
The point here is that because the Kerr metric $g^{ik}$ is known
explicitly in terms of $r$ and $\theta$, and because the
distribution of the angular momentum is given by the assumption
(\ref{distribution-momentum-entropy}), the right hand side of
(\ref{master}) is an explicitly known function of $r$ and
$\theta$. One may denote this function by $F(r, \theta)$, and
combine equations (\ref{equipressure-shape-first}) and
(\ref{master}) into the standard form of an ordinary differential
equation,
\begin{equation}
\label{differential} \frac{d\theta}{dr} = F(r, \theta),
\end{equation}
with the explicitly known right hand side. It may be directly
integrated to get all possible locations of of the isobaric
surfaces.

\citet{qia-et-al-2008} have recently argued that a physically
flexible and realistic angular momentum distribution may be
assumed of the form,
\begin{equation}
\label{distribution-recent} {\cal L}(r, \theta) = \left \{
\begin{array}{l}
{\cal L}_0\left( \frac{{\cal L}_K (r)}{{\cal L}_0}\right)^{\beta}
\,(\sin\theta)^{2\gamma} \quad {\rm for} ~~ r > r_{ms}~~
\\
\\
{\cal L}_{ms}\,(\sin\theta)^{2\gamma} \quad {\rm for} ~~ r <
r_{ms}~~
\end{array}
\right \},
\end{equation}
where ${\cal L}_{ms} \equiv {\cal L}_0\,[{\cal L}_K(r_{ms})/{\cal
L}_0]^{\beta}$, and ${\cal L}_0$, $\beta$, and $\gamma$ are the three free
constant parameters of the distribution.
\begin{figure}
\begin{center}
  \includegraphics[width=.24\textwidth]
  {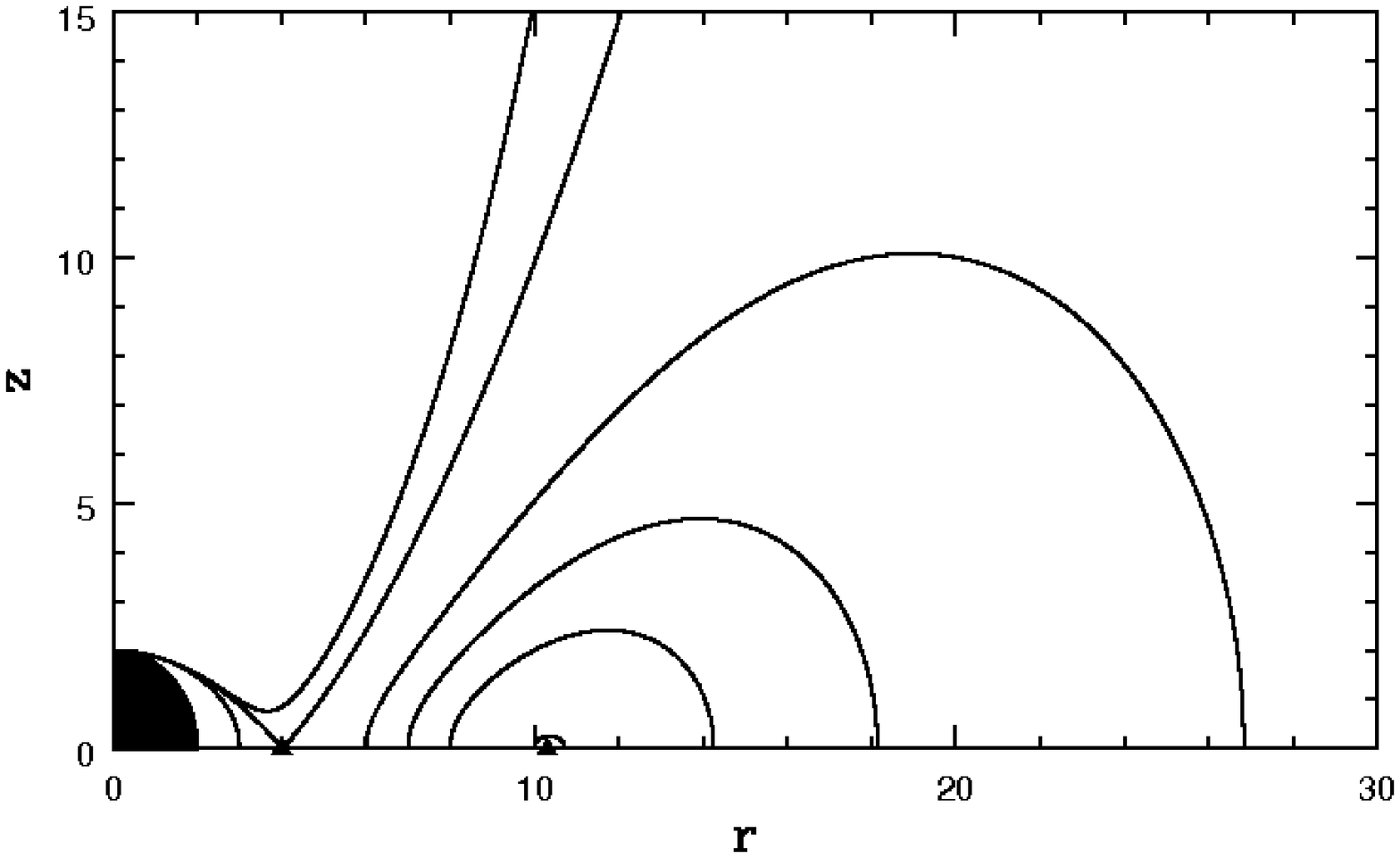}
  \includegraphics[width=.24\textwidth]
  {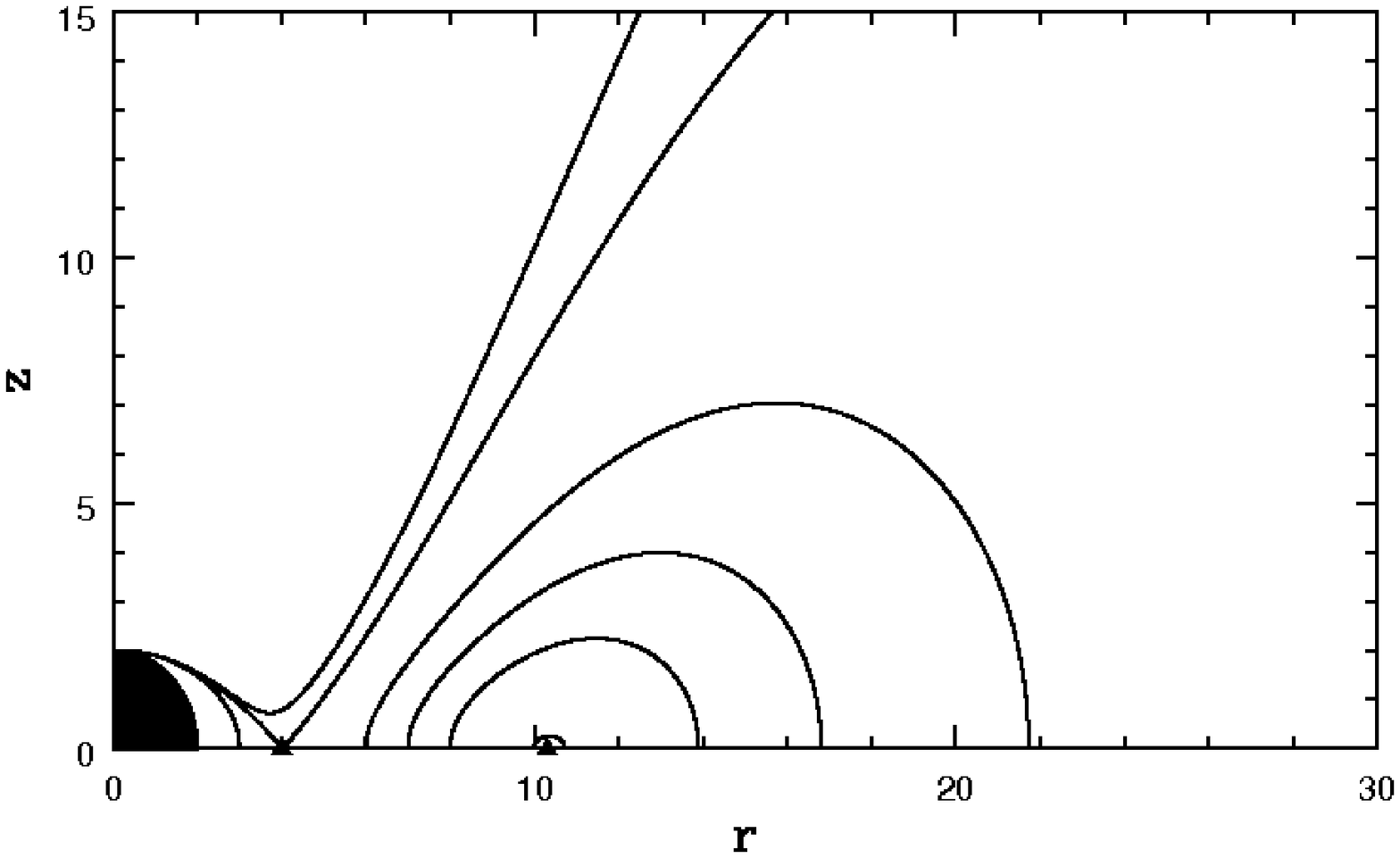}
  \includegraphics[width=.24\textwidth]
  {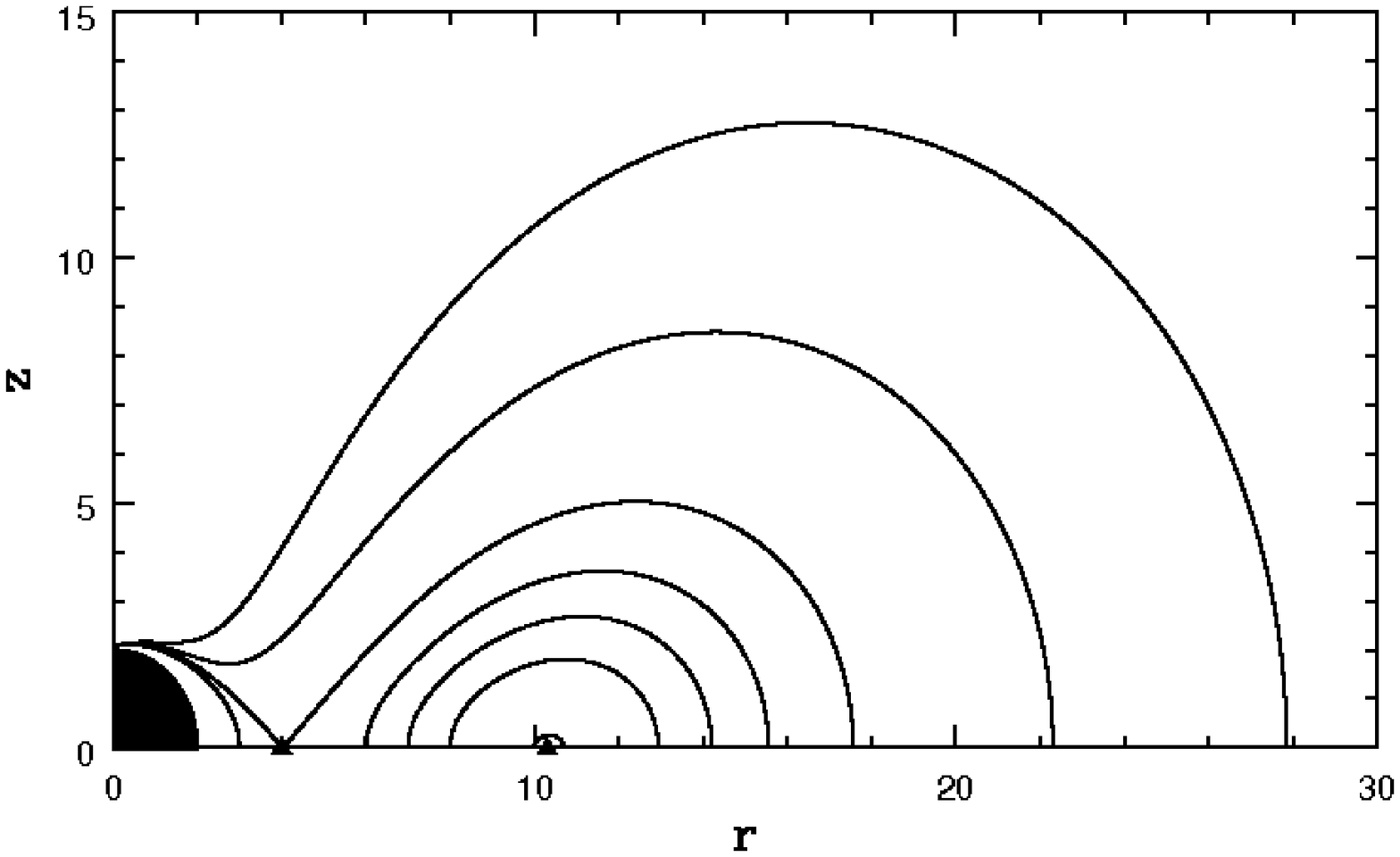}
  \includegraphics[width=.24\textwidth]
  {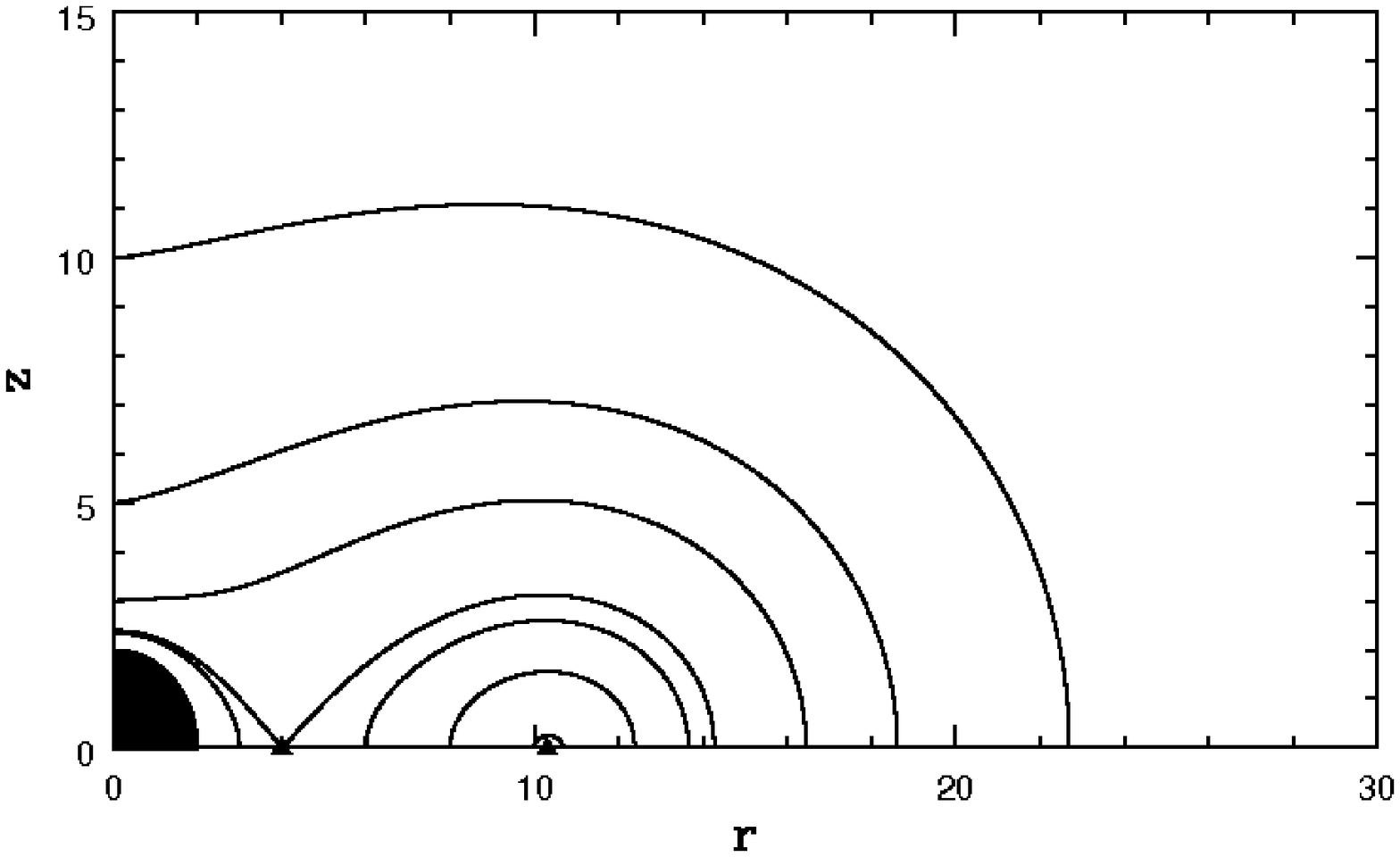}
\end{center}
\begin{center}
   \includegraphics[width=.24\textwidth]
   {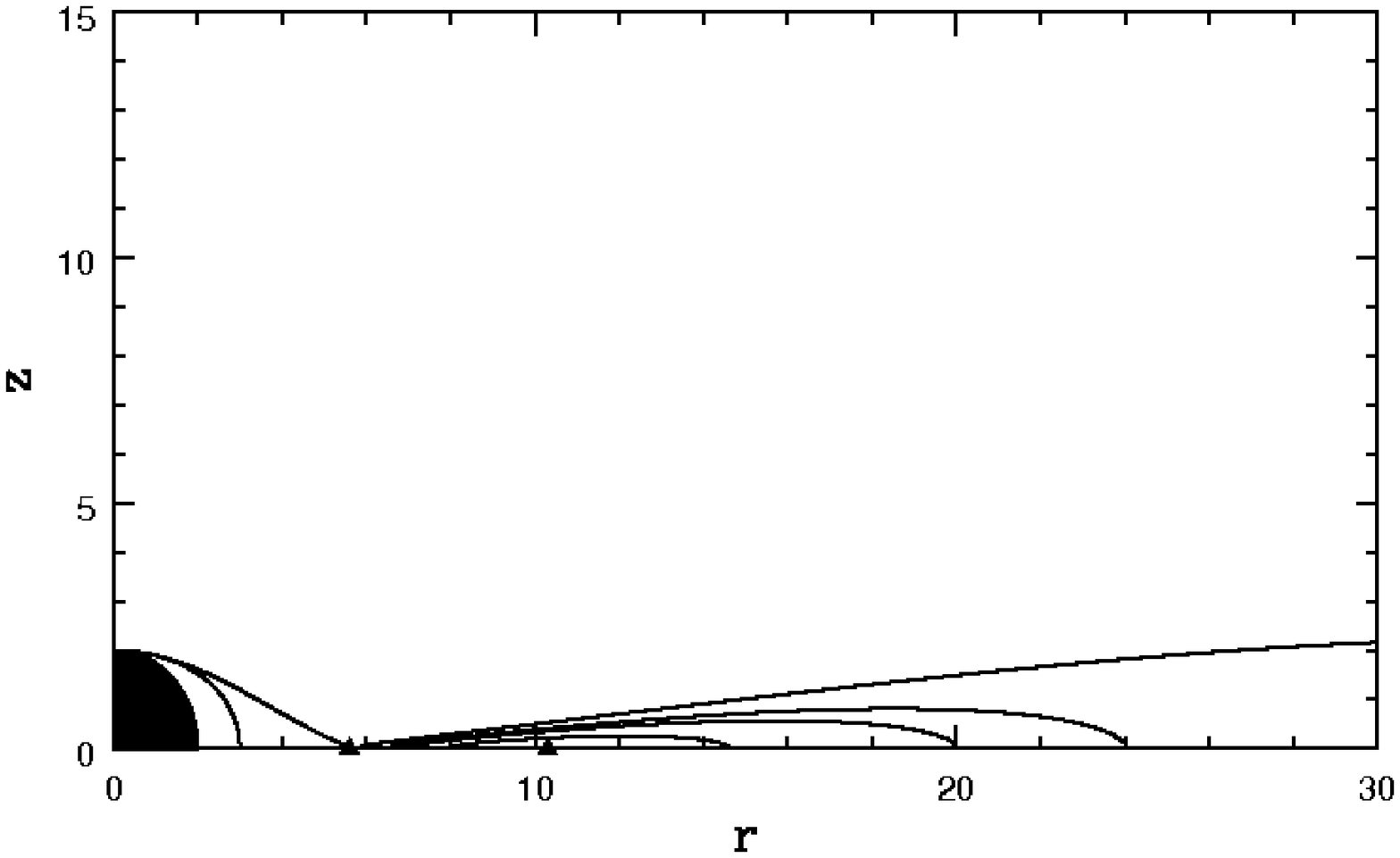}
   \includegraphics[width=.24\textwidth]
   {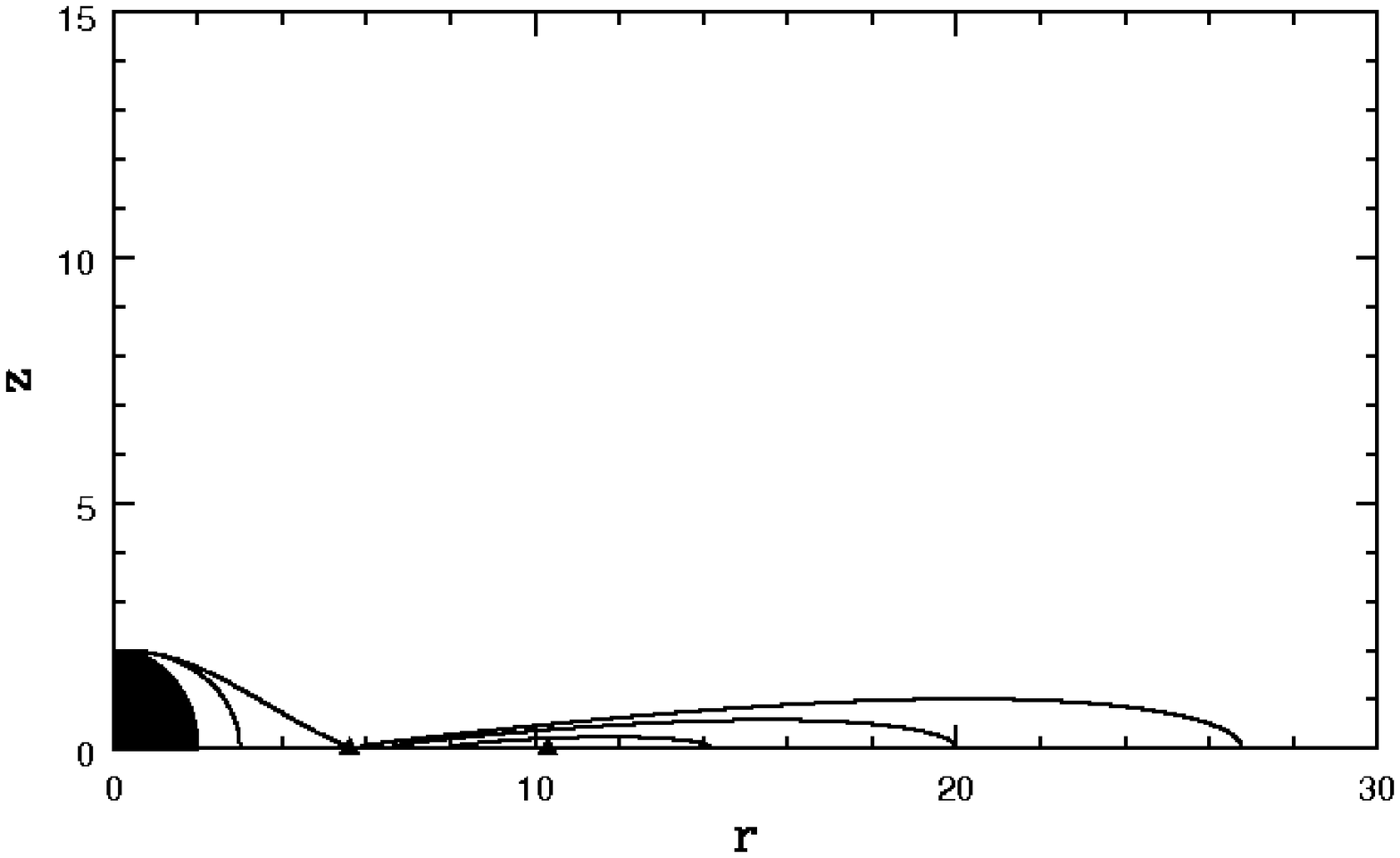}
   \includegraphics[width=.24\textwidth]
   {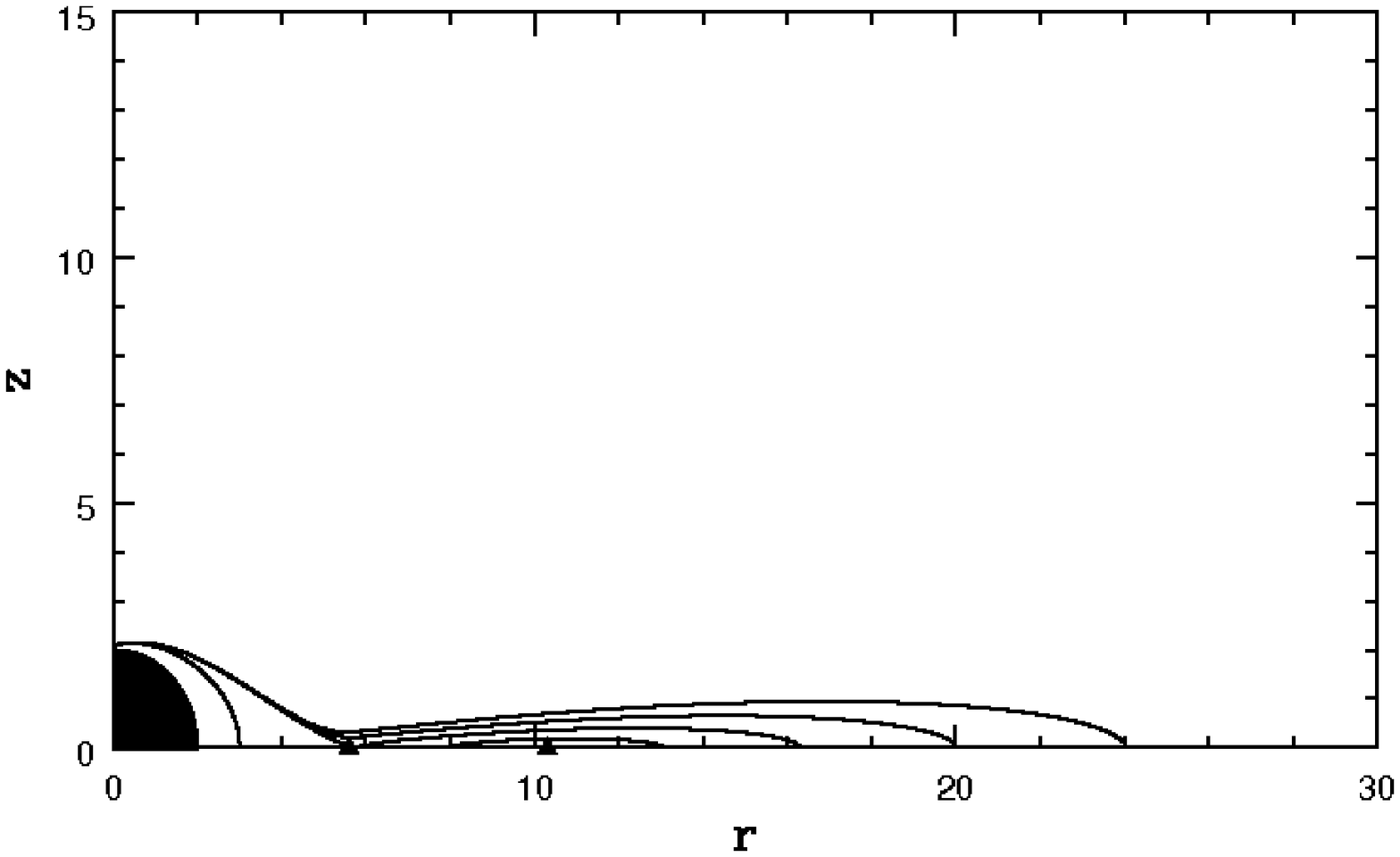}
   \includegraphics[width=.24\textwidth]
   {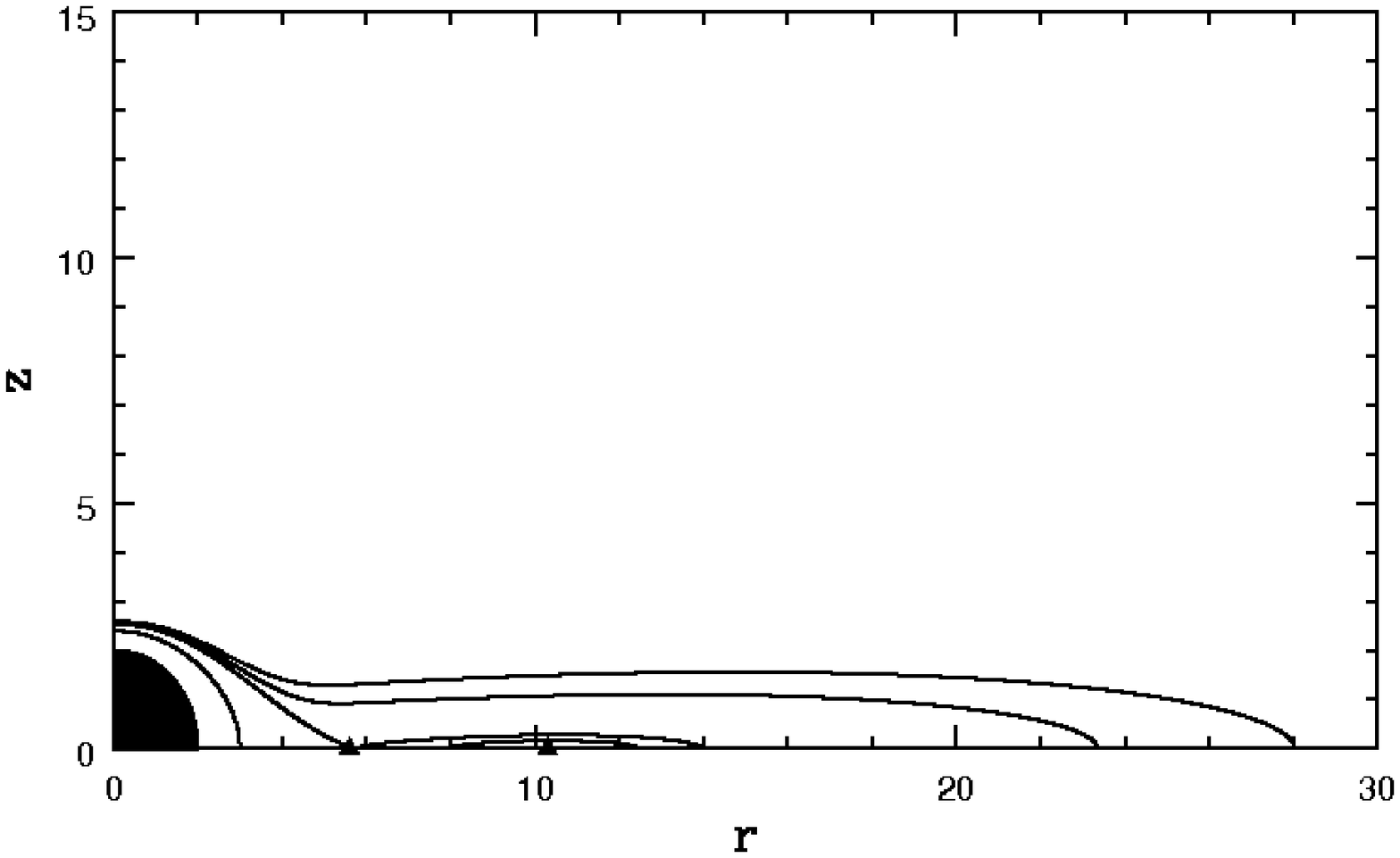}
\end{center}
   \caption{Shapes of Polish doughnuts. Upper row: for almost constant
   angular momentum, $\beta \approx 0$. Lower row: for almost Keplerian
   angular momentum, $\beta \approx 1$. The parameter $\gamma$ increases
   in equal steps from left, $\gamma = 0$, to right, $\gamma = 1$. }
   \label{qian-lei}
\end{figure}
Figure \ref{qian-lei} shows several models calculated for
different choices of ${\cal L}_0, \beta, \gamma$.

Recently \citet{kom-2006} constructed models of magnetized Polish
doughnuts.


\section{The Funnel} \label{section-funnel}


Long, narrow funnels along the axis are a genuine feature of
Polish doughnuts. Bohdan realized that they may be relevant for
beaming radiation to highly super-Eddington fluxes and
collimating relativistic jets.

\subsection{Super-Eddington luminosities in funnels}

In this section, I will use Newton's theory and cylindrical
coordinates $r, z, \phi$. In Newton's theory, the first integral
(\ref{first-integral}) that describes equilibrium of the constant
angular momentum, constant entropy fluid takes the form,
\begin{equation}
\label{equipotential-Newton} -{GM \over {\left( r^2 + z^2
\right)^{1\over 2}}} + {1\over 2} {{\cal L}^2 \over r^2} + W(p) =
{\rm const}.
\end{equation}
This equation cannot be obeyed at the rotation axis (where $z \not
= 0$, $r = 0$), which has an obvious, but important, consequence:
no isobaric surface may cross the axis. For a constant angular
momentum fluid, isobaric surfaces must be toroidal, or open.
The marginally open surface has just one point $(r=\infty, z=0)$
at infinity. This particular surface encloses the {\it largest
possible} torus. From (\ref{equipotential-Newton}) it is obvious
that in this case $W(p) = W(0) = 0$. The maximal pressure is located at a
circle $z=0, r=r_0 = GM/{\cal L}_0^2$. Using the radius $r_0$ as a
scale, $\xi = r/r_0$, $\eta = z/r_0$, $w = W/(GM/r_0)$, one may
write equation (\ref{equipotential-Newton}) in the dimensionless
form, and solve for $\eta = \eta (\xi, w)$ to obtain,
\begin{equation}
\label{equipotential-2} \eta^2 = Q^2(\xi) \equiv 4\xi^4 \left( 1 -
2 \xi^2 w \right)^{-2} - \xi^2,~~-1/2 \le w \le w_S.\,
\end{equation}
The value $w = -1/2\,$ gives the location of the center, and $w =
w_S \le 0\,$ the location of the surface. For the {\it slender
torus} $w_S \approx -1/2$, and the {\it fat torus} $w_S \approx
0$.
\begin{figure}[ht]
\begin{center}
\includegraphics[width=.5\textwidth]{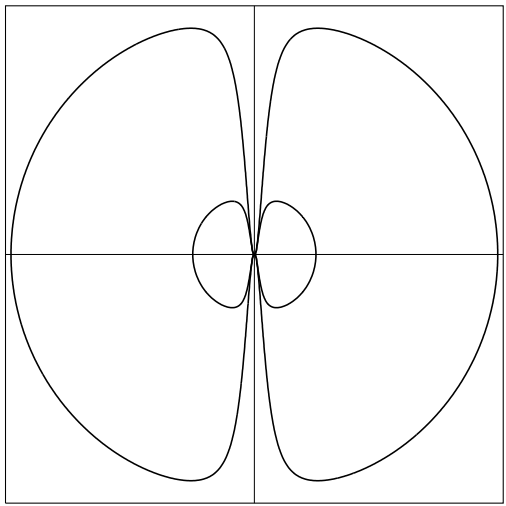}
\end{center}
\caption[]{Contours of equipotential surfaces on the meridional
cross section of a fat torus with constant angular momentum, i.e. with
$w_s \approx 0$. Close to axis the isobaric surfaces form a
pair of conical funnels.} \label{fat-torus}
\end{figure}
Very far from the axis of rotation one has $|2w\xi | \gg 1$.
Inserting this into (\ref{equipotential-2}) one gets $\eta^2 +
\xi^2 = 1/w^2$ which means that far from the rotation axis the
equipotential surfaces are spheres with radius $1/|w|$, and that
the outer radius of the torus is at $r_{\rm out} = r_0/|w_S|$.
Spherical equipotentials are in accord with the fact that very far
from the axis, $\xi \gg 1$, the centrifugal force $\sim {\cal
L}_0^2/\xi^3$ is negligible with respect to the gravitational
force $\sim GM/\xi^2$. Therefore, the effective gravity is
determined by Newton's attraction alone, as for spherical stars.
For the same reason, the radiation power from this part of the
surface of a fat, radiation-pressure-supported torus (i.e. Polish
doughnut) is {\it one Eddington luminosity}, the same as from a
spherical non-rotating, radiation-pressure-supported
star\footnote{Martin Rees told me as early as 1980 that to him this
implied that Polish doughnuts, with at least one Eddington
luminosity from whatever direction, were obviously too luminous to
describe very ``dim'' active galactic nuclei, such as radio
galaxies: ``while apparently supplying tremendous power to their
extended radio-emitting regions, the nuclei of most radio galaxies
emit little detectable radiation.'' He and his collaborators at
Cambridge later found a possible solution to this puzzle in terms
of the {\it ion-pressure-supported tori} \cite{ree-et-al-1982}.
The ion tori have the same shapes as Polish doughnuts (in
particular funnels) but have much lower, indeed very sub-Eddington,
luminosities. The power in jets comes from tapping the rotational
energy of the central black hole by the Blandford-Znajek mechanism
\cite{bla-zna-1977}.}. Note, however, that the asymptotically
spherical shape of a fat torus is a direct consequence of the
assumption ${\cal L}(r,z) ={\rm const}$, which was made ad hoc. If
one adopts a more physically realistic assumption that
asymptotically ${\cal L}(r,z)={\cal L}_K$, as in our model
(\ref{distribution-recent}), one may use the standard
Shakura-Sunyaev model in its radiation pressure version to get
the asymptotic shape of the fat torus,
\begin{equation}
\label{asymptotic-thickness} z_{\infty} = {{3\sigma_T} \over {8\pi
c m_p}}\, {\dot M}.
\end{equation}
Closer to the axis, $|2w\xi | \sim 1$, which means that $\xi^2
\sim -1/2w$, and this together with (\ref{equipotential-2}) gives
$\eta^2/\xi^2 = (1/|w|) - 1 = (1/\sin^2 \theta) - 1$, i.e. that
closer to the axis, equipotential surfaces corresponding to $w
\sim w_S \sim 0$ have conical shapes with the half opening angle
$\theta \sim \sqrt w$. The surfaces are highly non-spherical because
centrifugal force dominates. Integrating effective gravity along
the conical funnel is elementary, and one gets \citep{jar-1980}
that $L^{\rm rot}_{\rm Edd}/L_{\rm Edd} = (1/2) \ln (1/|w_S|)$.
This estimate may be used to find the total luminosity for the
radiation-pressure-supported fat torus. i.e. a Polish doughnut,
\begin{equation}
\label{logarithm} {L_{\rm total} \over L_{\rm Edd}} \approx
{L_{\rm Edd}^{\rm rot} \over L_{\rm Edd}} \approx {1\over 2} \ln
\left({r_{\rm out}\over r_{0}}\right ) = 1.15\,\log\left ({r_{\rm
out} \over r_0}\right ).
\end{equation}
It should be clear from our derivation that the logarithmic
scaling of the luminosity with the torus size is a genuine
property of Polish doughnuts, including those that have
an angular momentum distribution that is not constant. The logarithm in
(\ref{logarithm}) is of crucial importance, as it prevents
astrophysically realistic doughnuts (i.e. with $r_{\rm out}/r_{0}
< 10^6$, say) from having highly super-Eddington luminosities. Thus,
the theory predicts that for such ``realistic'' fat tori, only
slightly super-Eddington total (isotropic) luminosities, $L_{\rm
total} \ge 7\,L_{\rm Edd}$, may be expected.

However, because the funnels have solid angles  $\theta^2 \sim
r_{0}/r_{\rm out }$, radiation in the funnels may be, in
principle, collimated to highly super-Eddington values $L_{\rm
coll} /L_{\rm Edd} = \Theta \sim r_{\rm out}/r_{0} \gg 1$. This
simple estimate agrees with a more detailed modelling of the
radiation field of the Polish doughnuts by \citet{sik-1981} and
\citet{mad-1988} who obtained $\Theta \le 10^2$ for disks with
$r_{\rm out}/r_{0} \sim 10^2$. A typical value that follows from
observational estimates for non-blazar active galactic nuclei
\citep[e.g.][]{mal-1989, cze-elv-1987} is $\Theta \sim 10$, but of
course for blazars and other similar sources, e.g. for ULXs, if
they are powered by stellar mass black holes, as argued by
\citet{kin-2002}, it must be $\Theta \gg 10$.

Such high values of $\Theta$ are also consistent with the idea,
suggested by \citet{pac-1980} and independently by
\citet{bel-1982}, that relativistic electron-positron $e^-e^+$
jets may be very effectively accelerated by the radiation pressure
in the fat tori funnels\footnote{Lynden-Bell called this an
``entropy fountain''.}. Note that if the flux in the funnel is
$\Theta$ times the Eddington flux, the $e^-e^+$ plasma feels the
``effective'' radiative force corresponding to the Eddington ratio
$m_p/m_e \approx 10^3$ times greater. Detailed calculations
\citep[e.g.][]{abr-pir-1980, sik-wil-1981, abr-ell-lan-1990}
demonstrate that indeed the $e^-e^+$ jets may be accelerated in
funnels up to the Lorentz factor $\gamma \le 5$. However, if jets
are initially pre-accelerated by some black-hole electrodynamical
processes \citep[such as the Blandford-Znajek
mechanism,][]{bla-zna-1977} to highly relativistic velocities
$\gamma > 10^6$, they will be decelerated in the funnels by
Compton drag, reaching the asymptotic Lorentz factor
\citep{abr-ell-lan-1990},
\begin{equation}
\label{lanza} \gamma_{\infty} = \left ( \Theta \,{m_p \over m_e}
\right )^{1\over 3} = 10 \times\Theta^{1\over 3}.
\end{equation}
Observations show that $\gamma_{\infty} < 10^2$, and thus equation
(\ref{lanza}) suggests that $\Theta < 10^3$.

\subsection{The physical reason for super-Eddington luminosities}

I will now review a very general argument, first presented by
\citet{abr-cal-nob-1980}, which shows that the upper limit for the
luminosity of rotating bodies in equilibrium must differ from the
Eddington limit. I will consider two ``astrophysical'' cases:

{\it Rotating stars.} The surface of the star has the topology of
a sphere. The whole mass $M$ is included in the sphere.

{\it Accretion disks.} The surface of the disk has the topology of
a torus. The mass of the disk $M_{\rm disk}$ is contained in the
torus, but the mass $M_{\rm centr}$ of the central black hole is
outside. The total mass $M = M_{\rm centr} + M_{\rm disk} \approx
M_{\rm centr}$, because $M_{\rm centr} \gg M_{\rm disk}$. In
accretion disk theory it is customary to neglect the mass of the
disk, so formally $M = M_{\rm centr}$, and $M_{\rm disk} =0$. The
Eddington limit always refers to the total mass $M$.

Let ${\bf f}_{\rm rad}$ be the local flux of radiation somewhere
at the surface of the body, $S$. The corresponding radiative force
is ${\bf F}_{\rm rad} = (\sigma_{\rm rad}/c){\bf f}_{\rm rad}$.
Let ${\bf F}_{\rm eff} = {\bf F}_{\rm grav} + {\bf F}_{\rm rot}$,
be the effective gravitational force, with ${\bf F}_{\rm grav} = m
\bnabla \Phi$ being the gravitational force ($\Phi$ is the
gravitational potential), and with ${\bf F}_{\rm rot} = m
(\Omega^2 r)\,{\bf e}_{\rm r}$ being the centrifugal force
($\Omega$ is the angular velocity, $r$ is the distance from the
axis of rotation, and ${\bf e}_{\rm r} = \bnabla r$ a unit vector
in the off-axis direction).

The necessary condition for equilibrium is ${\bf F}_{\rm rad} <
{\bf F}_{\rm eff}$. From this one deduces the Eddington limit for
rotating perfect-fluid bodies,
\begin{equation}
L = \int_{\rm S} {\bf f}_{\rm rad}\cdot d{\bf S} = {c \over
\sigma_{\rm rad}}\int_{\rm S} {\bf F}_{\rm rad}\cdot d{\bf S}
\,<\, {c \over \sigma_{\rm rad}}\int_{\rm S} {\bf F}_{\rm
eff}\cdot d{\bf S} \equiv L_{\rm Edd}^{\rm rot}
\end{equation}
Using Gauss's theorem to transform the surface integral of ${\bf
F}_{\rm eff}$ into a volume integral of $\bnabla \cdot {\bf
F}_{\rm eff}$, Poisson's equation $\bnabla^2 \Phi = 4\pi G\rho\,$,
and introducing the specific angular momentum $\ell = \Omega r^2$,
one gets after twenty or so lines of simple algebra,
\begin{equation}
L_{\rm Edd}^{\rm rot} = L_{\rm Edd} \left [\,X^2_{\rm mass} +
X^2_{\rm shear} - X^2_{\rm vorticity}\,\right ],
\end{equation}
where $L_{\rm Edd}$ is the standard Eddington limit for a
non-rotating star,
\begin{equation}
\label{Eddington-standard} L_{\rm Edd} = 1.4 \times 10^{38} \left
({M \over M_{\odot}}\right) [\,{\rm erg}\,\,{\rm sec}^{-1}]\,.
\end{equation}
and where the dimensionless number $X^2_{\rm mass}$ depends on
whether the body is a star, or an accretion disk,
\begin{equation}
X^2_{\rm mass} = {1 \over M } \int_{\rm V} \rho dV = \left \{
\begin{array}{l}
1 \quad {\rm for~stars,}
\\
\\
0 \quad {\rm for~accretion~disks,}
\end{array}
\right \}.
\end{equation}
Here $X^2_{\rm shear}$ and $X^2_{\rm vorticity}$ are dimensionless,
necessarily positive quantities, equal to shear and vorticity
integrated over the whole volume of the body,
\begin{equation}
X^2_{\rm shear} = {1\over {16\pi G M}}\int_{\rm V} r^2 ({\bnabla}
\Omega \cdot \bnabla \Omega)\,dV,
\end{equation}
\begin{equation}
X^2_{\rm vorticity} = {1\over {16\pi G M}}\int_{\rm V} r^{-2}
(\bnabla {\cal L} \cdot \bnabla {\cal L})\,dV.
\end{equation}
Shear increases the Eddington limit, and vorticity decreases it.

The rotation of astrophysical objects is far from simple, but
insight can be gained by considering a simple power law for the
angular momentum distribution, ${\cal L}(r, z) = {\cal L}_0 r^a$,
with ${\cal L}_0$ and $a$ constant. Rigid rotation has $a=2$, Keplerian
rotation $a=1/2$, and constant angular momentum rotation $a=0$.
$X_{\rm shear}$ and $X_{\rm vorticity}$ are related to $a$ by
$X^2_{\rm shear}/X^2_{\rm vorticity} =(a-3)^2/a^2$.
This means that $X^2_{\rm shear} > X^2_{\rm vorticity}$ when $a <
3/2$.

Rotating stars have $X^2_{\rm shear} < X^2_{\rm vorticity}$,
because they rotate almost rigidly. Thus, for rotating, radiation
pressure supported stars, $L \approx L_{\rm Edd}^{\rm rot} <
L_{\rm Edd}$ {\it always}. Contrary to this,
tori with constant angular momentum are dominated by shear,
$X^2_{\rm shear} \gg 1 \gg
X^2_{\rm vorticity}$ and consequently, when they are radiation
pressure supported, $L \approx L_{\rm Edd}^{\rm rot} \gg L_{\rm
Edd}$.

\subsection{Rise and Fall of the Polish Doughnuts}

At the time of their d{\'e}but, Polish doughnuts could
theoretically confirm the observed super-Eddington luminosities,
highly collimated beams of radiation, and perhaps even the
relativistic speeds of jets, fulfilling the principle attributed
to Eddington: {\it one should never believe any experiment until
it has been confirmed by theory.} These virtues initially attracted
some interest in the astrophysical community, but the
interest quickly drained with the discovery of the
Papaloizou-Pringle instability. It was thought that Polish
doughnuts must necessarily suffer from the instability and thus,
in reality, they cannot exist. The important discovery by
\citet{bla-1987} that the Roche lobe overflow stabilizes Polish
doughnuts against the Papaloizou-Pringle instability came too
late --- in the advent of numerical simulations of black hole
accretion flows. Too late, because numerical simulations
rediscovered and absorbed many of the results of Polish doughnuts.
Today, these results exist in the consciousness of many
astrophysicists as a set of several numerically established,
important but unrelated facts. They do not form a consistent
scheme that the Polish doughnuts once offered: clear, simple,
following directly from the black hole physics\footnote{There are
at least three brilliant and very pedagogical expositions of the
Polish doughnuts scheme: two by Paczy{\'n}ski himself
\citep{pac-1982, pac-1998} and one in the text book by
\citet{kin-fra-rai-2002}.}.


\section{The Roche Lobe} \label{section-roche}


A powerful analytic model for accretion flows near the ISCO was worked
out in terms of the flow equipotential structure by Paczy{\'n}ski
and collaborators thirty years ago in Warsaw \citep[see
e.g.][]{abr-1978, koz-1978, jar-1980, abr-1981}. The Warsaw model
accurately describes the flow hydrodynamics for the standard
Shakura-Sunyaev thin disks, slim disks, adafs, ion tori, and thick
Polish doughnuts. It impressively agrees with more recent
numerical simulations of accretion flows, as directly checked e.g.
by \citet{igu-abr-2000}.

\begin{figure}
  \begin{center}
  \includegraphics[width=0.45\textwidth]{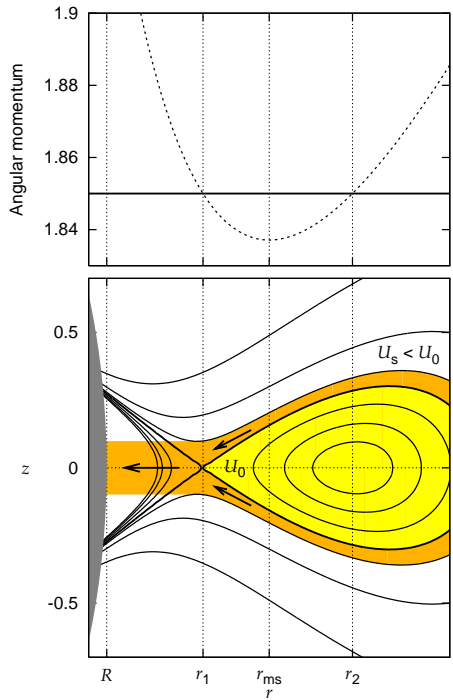}
  \caption{Upper figure shows the Keplerian angular momentum (the dotted
  line) and the angular
  momentum in the flow (the solid line). Lower figure shows the equipotential
   surfaces (solid
  lines) and distribution of fluid (shade).}
  \label{fig01}
  \end{center}
\end{figure}

In an equilibrium described by the Bernoulli equation
(\ref{equipotential-Newton}), surfaces of constant enthalpy,
pressure and density coincide with surfaces of constant effective
potential, $\mathcal{U}(r,z) = \mathrm{const}$. The surface of the
disk is given by $P = \mathrm{const} = 0.\,$ Thus, equilibria may
only exist if the disk surface corresponds to one of the
equipotentials inside the Roche lobe, i.e. in the region indicated
by yellow in Figure \ref{fig01}.  If the fluid distribution
overflows the Roche lobe, i.e. the surface of the disk for $r \gg
r_1$ coincides with $\mathcal{U}(r,z) = \mathcal{U}_S <
\mathcal{U}_0,\,$, the equilibrium in the region $r \le r_1$ is
not possible, and the disk will suffer a dynamical mass loss, with
accretion rate ${\dot M}_{\rm in}$.

\subsection{The stationary Roche lobe overflow}

\begin{figure}
  \begin{center}
  \includegraphics[width=0.7\textwidth]{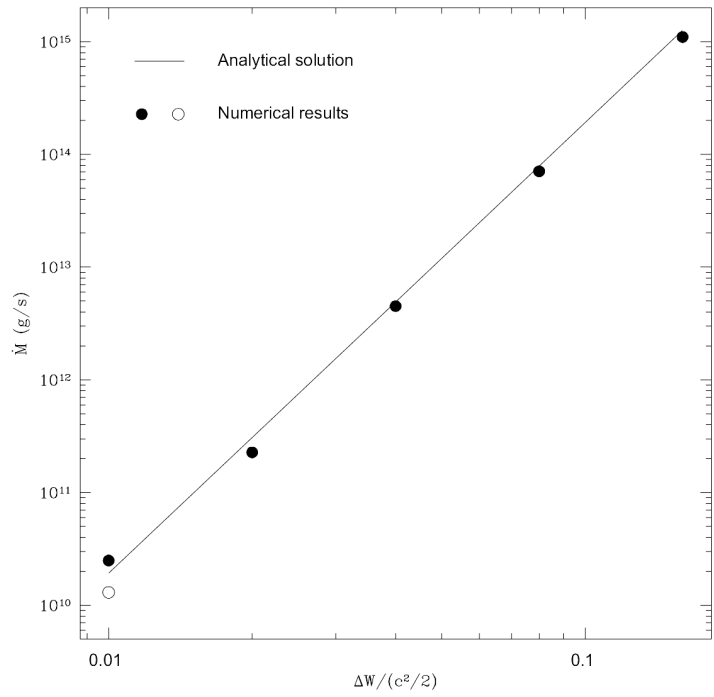}
  \caption{The line is according to the analytic formula (\ref{jedrzejec-formula})
  and the points are from non-stationary, 3D numerical simulations by
  \cite{igu-abr-2000}.}
  \end{center}
  \label{igor-agreement}
\end{figure}

An analytic formula for ${\dot M}$ for stationary flows was first
calculated by \cite{koz-1978}, who used Einstein's theory. Here
we review another derivation, done by \cite{abr-1985}, as we later
use the same assumptions and notation to calculate ${\dot M}$ for
a non-stationary (oscillating) disk. In particular, we assume that
the Roche lobe overflow is small (quadratic in disk thickness
$H$),
\begin{equation}
  h(r_1,z)\,=\,h^\star-\frac{1}{2}\kappa^2 z^2,
  \quad
  \kappa^2\equiv-\left(\frac{\partial^2 h}{\partial z^2}\right)_{\!L},
  \quad
  H=\frac{\sqrt{2h^\star}}{\kappa},
  \label{eq:quadrh}
\end{equation}
that the equation of state is polytropic, and that the radial
velocity $v^r$ connected to the mass loss through the cusp equals
the sound speed $c_\mathrm{s}$,
\begin{equation}
  P=K\rho^{1+1/n}, \quad
  v^z \ll v^r\,= \,c_\mathrm{s}=\sqrt{\frac{h}{n}}.
\end{equation}
In the equations above $h^\star=h(r_1,0)$ denotes a maximal
value of the enthalpy on the cylinder $r=r_1$ and the subscript $L$
stands for the evaluation of the derivative at the point
$[r_1,0]$. The local mass flux $\dot{m}=\rho v^r =
{h^{n+1/2}}/{K^n(1+n)^n n^{1/2}}$, vertically integrated through
the cusp thickness and azimuthally around, gives the desired total
mass flux in terms of the enthalpy,
\begin{eqnarray}
  \dot{M} &=&
  \int_0^{2\pi} r_1 d\varphi \int_{-H}^{+H}\dot{m}dz=\nonumber\\
  &=&(2\pi)^{3/2}\frac{r_1}{\kappa n^{1/2}}\left[\frac{1}{K(n+1)}\right]^n
  \frac{\Gamma(n+3/2)}{\Gamma(n+2)}h_0^{n+1}.
  \label{eqn07}
\end{eqnarray}
Here $\Gamma(x)$ is the Euler gamma function. From
${v^2}/{2}+h+\mathcal{U}=\mathcal{U}_\mathrm{S}\,$ one gets
$\Delta\mathcal{U}= \mathcal{U}_\mathrm{S} - \mathcal{U} =
\left(1+{1}/{2n}\right)h\,$ and from
\begin{equation}
  \kappa^2\equiv-\left(\frac{\partial^2 h}{\partial z^2}\right)_L
  \left(\frac{n}{n+1/2}\right)\omega_z^2,
  \quad
  \omega_z^2 \equiv
  \left(\frac{\partial^2 \mathcal{U}}{\partial z^2}\right)_L,
  \label{eq:kappa}
\end{equation}
we recover the result obtained by \cite{abr-1985},
\begin{equation}
  \dot{M} = A(n)\frac{r_1}{\omega_z}\Delta\mathcal{U}^{n+1},
 \label{jedrzejec-formula}
\end{equation}
\begin{equation}
  A(n)\equiv(2\pi)^{3/2}\left[\frac{1}{K(n+1)}\right]^n
  \left[\frac{1}{n+1/2}\right]^{n+1/2}
  \frac{\Gamma(n+3/2)}{\Gamma(n+2)}.
\end{equation}
Figure \ref{igor-agreement} shows an excellent agreement between
the analytic formula (\ref{jedrzejec-formula}) and results of
large-scale, 3D, non-stationary numerical simulations.

\subsection{Non-stationary Roche lobe overflow}

We will calculate $\dot{M}$ for non-stationary oscillating flows
in a special but important case, assuming that the poloidal part
of velocity $\vec{v} = (v^r, v^z)$ may be derived from a
potential. In this case, the non-stationary version of the
Bernoulli equation has the form,
\begin{equation}
  \frac{\partial\chi}{\partial t}+\frac{v^2}{2}+h+\mathcal{U}
  \mathcal{U}_\mathrm{S}, \quad {\rm with} \quad \vec{v} = \nabla\chi.
  \label{eqn11}
\end{equation}
We assume that the oscillation is a small, non-stationary perturbation
to the stationary flow considered in the previous section,
\begin{equation}
  \chi(\vec{r},t)=\chi_{(0)}(\vec{r})+\epsilon\chi_{(1)}(\vec{r},t),
\end{equation}
where the subscript (0) refers to the stationary flow and the
dimensionless  parameter $\epsilon\ll1$ characterizes the strength of
the perturbation. From the definition $\vec{v} = \nabla\chi$ one
derives,
\begin{eqnarray}
  v^2 &=& v_{(0)}^2+2\epsilon\vec{v}_{(0)}\cdot\vec{v}_{(1)}
  +\epsilon^2 v_{(1)}^2
  \nonumber \\
  &=& c_\mathrm{s}^2+2\epsilon c_\mathrm{s}
  \frac{\partial\chi_{(1)}}{\partial r}+
  \epsilon^2\left[
  \left(\frac{\partial\chi_{(1)}}{\partial r}\right)^{\!2}\!\!\!+\!
  \left(\frac{\partial\chi_{(1)}}{\partial z}\right)^{\!2}
  \right].
  \label{eq:v2}
\end{eqnarray}
The perturbed enthalpy profile can be approximated by an expansion
\begin{equation}
  h = h_{(0)} + \epsilon h_{(1)} + \epsilon^2 h_{(2)} +
  \mathcal{O}(\epsilon^3).
  \label{eq:hexp}
\end{equation}
By substituting equations (\ref{eq:v2}) and (\ref{eq:hexp}) into
the Bernoulli equation (\ref{eqn11}) and equating coefficients of
the same powers of $\epsilon$, we get
\begin{eqnarray}
  h_{(1)}&=&-\frac{\partial\chi_{(1)}}{\partial t}-
  \left({\frac{\bar{h}}{n}}\right)^{1/2}
  \frac{\partial\chi_{(1)}}{\partial r},
  \label{eq:h1} \\
  h_{(2)}&=&-\frac{1}{2}\left[
  \left(\frac{\partial\chi_{(1)}}{\partial r}\right)^{\!2}\!\!+
  \left(\frac{\partial\chi_{(1)}}{\partial z}\right)^{\!2}\right].
  \label{eq:h2}
\end{eqnarray}
This way, we express perturbations of fluid quantities in terms
of a perturbation $\chi_{(1)}$. To progress further, one must know
the function $\chi_{(1)}=\chi_{(1)}(t,r,z)$ that describes
oscillations. Finding it in general is a difficult global problem.
We do not attempt to solve it here. Instead, we describe
oscillations by an ansatz,
\begin{equation}
  \chi_{(1)}=z v_z\cos\omega t, \quad
  v_z = \mathrm{const}.
  \label{eqn13}
\end{equation}
which models a vertical ``epicyclic'' oscillation with frequency
$\omega$ that rigidly moves the fluid up and down across the
equatorial symmetry plane. Such a rigid mode, with the
eigenfrequency equal to the epicyclic vertical frequency, $\omega
= \omega_z$, has been recently identified by \citet{abr-2006} as a
combination of eigenmodes of slender torus oscillations found
previously by \cite{bla-1985}. This classic paper gives {\it all}
possible modes, i.e. their eigenfrequencies and eigenfuncions, in
terms of exact, simple, and explicit analytic formulae.

\begin{figure*}
  \begin{center}
    \resizebox{\hsize}{!}{
    \includegraphics[width=0.49\textwidth]{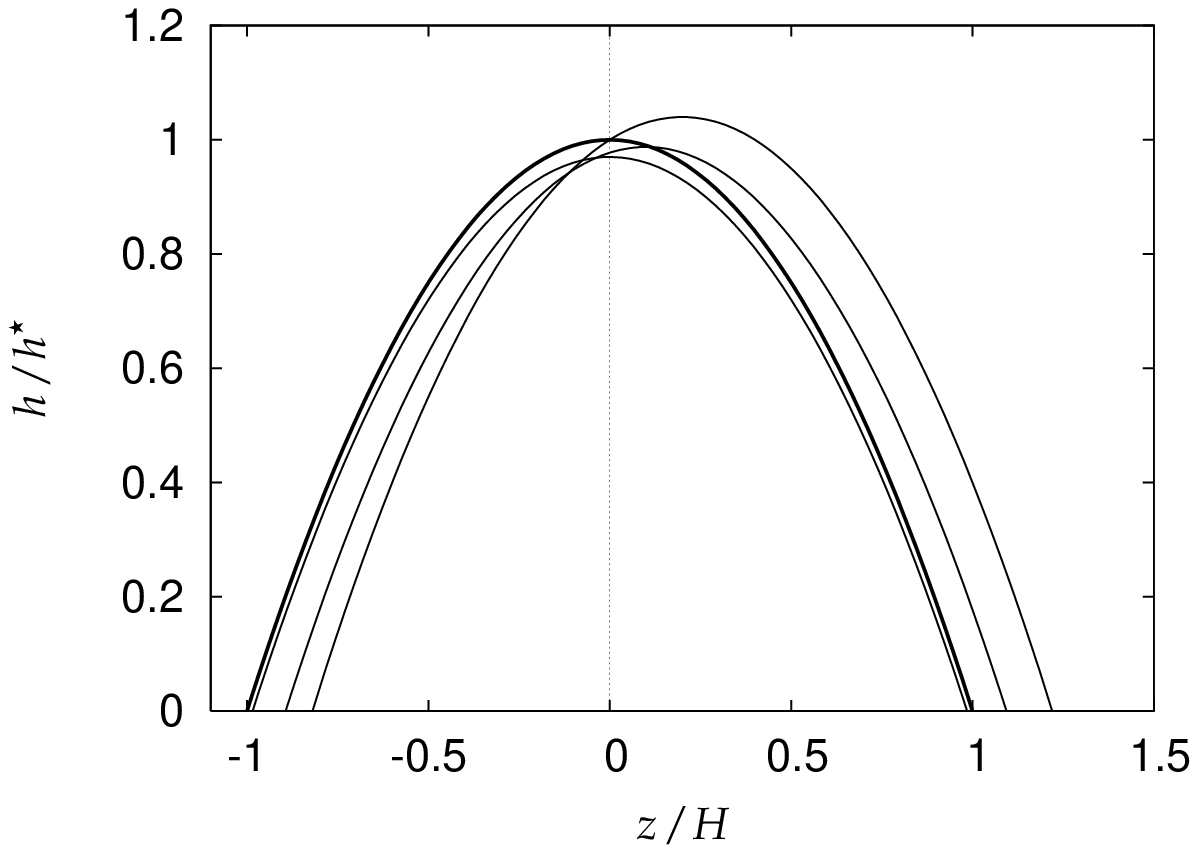}
    \hfill
    \includegraphics[width=0.49\textwidth]{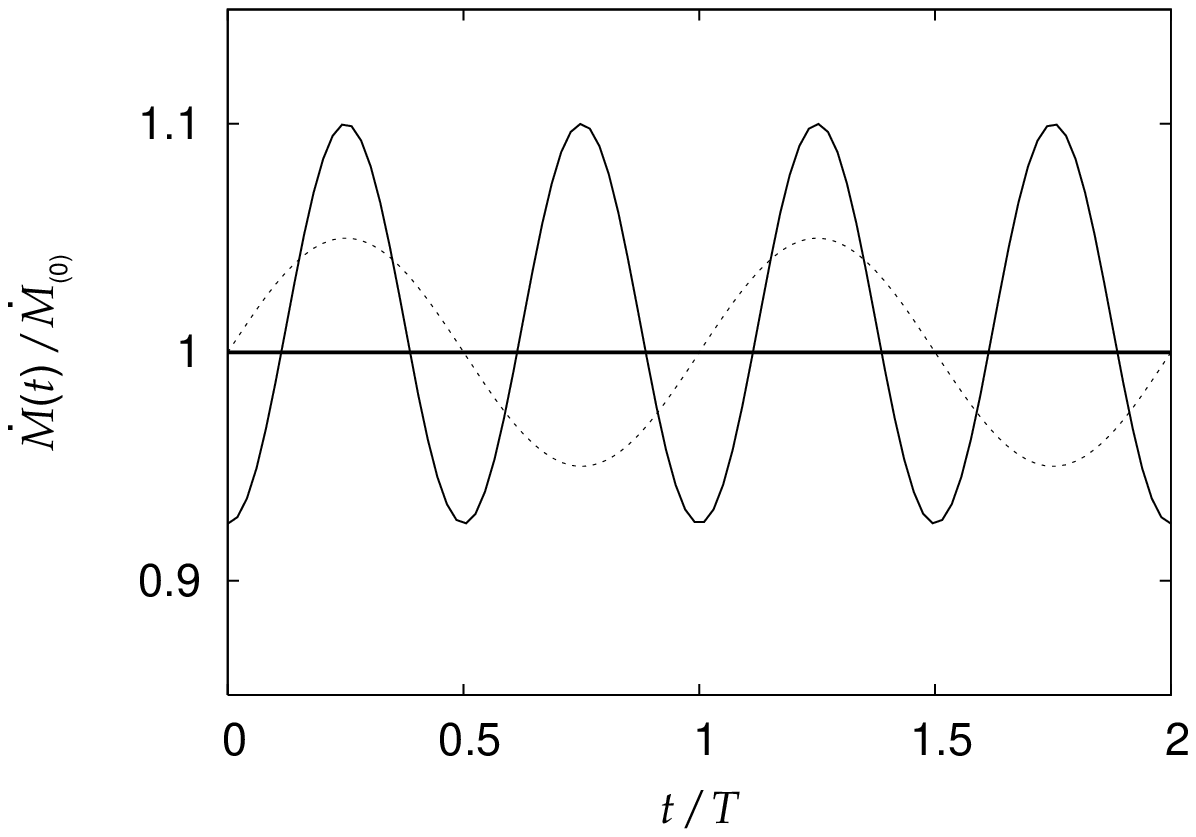}
    }
  \end{center}
  \caption{Left: The vertical profiles of the enthalpy $h(r_1,z)$ on the
  cylinder $r=r_1$ during vertical disk oscillations (thin lines). The
  amplitude of oscillations is $\delta Z=0.2 H$ and we chose the
  polytropic  index of the fluid $n=1.5$. The figure captures profiles
  with enthalpy maxima at $\delta z=0,0.1 H$  and $0.2 H$. The
  enthalpy profile for the unperturbed stationary disk is also shown
  (thick line). Right: The modulated accretion rate from the oscillating
  disk (thin solid line). The accretion rate for the stationary disk  is
  plotted by the thick line. Time is rescaled by the period of oscillations.
  For reference we plot also the phase of disk oscillations (dotted line)
  The accretion rate is modulated with twice the frequency  of
  oscillations.}
  \label{fig:profiles}
\end{figure*}

The quantity $\epsilon v_z=\mathrm{const}$ can be interpreted as
the amplitude of the vertical velocity. Equations (\ref{eq:h1})
and (\ref{eq:h2}) give
\begin{equation}
  h_{(1)}=z v_z \omega \sin\omega t, \quad
  h_{(2)}=-\frac{1}{2} v_z^2 \cos^2\omega t.
\end{equation}
The vertical profile of the enthalpy at $r=r_1$ is
\begin{eqnarray}
  h(r_1,z,t)\!\!&=&\!\!h^\star-\kappa^2 z^2 + \epsilon z v_z\omega\sin\omega t-
  \nonumber\\
  \quad
  \!\!&-&\!\!
  \frac{1}{2}\epsilon^2 v_z^2\cos^2\omega t +
  \mathcal{O}(\epsilon^3),
\end{eqnarray}
which is quadratic in the variable $z$ (see the left panel of
Fig.~\ref{fig:profiles}). The position of the enthalpy maximum on
the cylinder  $r=r_1$ is shifted from $z=0$ to height $\delta
z(t)$, given as
\begin{equation}
  \delta z(t)= \delta Z \sin\omega t, \quad
  \delta Z = \epsilon\,\frac{\omega v_z}{\kappa^2}.
\end{equation}
We can interpret $\delta Z$ as the amplitude of the oscillations.
Also the value of enthalpy in the maximum differs from the
stationary case by
\begin{eqnarray}
  \delta h^\star\!\!&\equiv&\!\!h(r_1,\delta z)-h^\star
  \nonumber\\
  \!\!&=&\!\!
  \frac{1}{2}\kappa^2\left[\delta z^2-\frac{\kappa^2}{\omega^2}
  (\delta Z^2-\delta z^2)\right]+\mathcal{O}(\epsilon^3).
  \label{eq:dh}
\end{eqnarray}
According to equation (\ref{eqn07}) the instantaneous accretion
rate depends on the maximal enthalpy as  $\dot{M}\propto
(h^\star)^{n+1}$. This relation can also be applied in the case of
vertical oscillations because the $z$-dependence of  enthalpy on
the cylinder $r=r_1$ remains quadratic and the
oscillations  do not contribute to the radial velocity of accreted
matter. Hence, using equations (\ref{eq:quadrh}), (\ref{eq:kappa})
and (\ref{eq:dh}) and assuming that the frequency of oscillations
equals the local vertical epicyclic frequency,
$\omega=\omega_z$, we arrive at our final result
\begin{eqnarray}
  \frac{\delta\dot{M}}{\,\,\,\dot{M}_{(0)}}\!\!\!&=&\!\!\!(n+1)
  \frac{\delta h^\star}{h^\star}
  \nonumber\\
  \!\!\!&=&\!\!\!
  \frac{2-p}{2-2p}\left[(1+p)\frac{\delta z^2}{H^2}-p\frac{\delta Z^2}{H^2}\right],
\end{eqnarray}
where $\delta \dot{M}\equiv\dot{M}-\dot{M}_{(0)}$ and $p\equiv
n/(n+1/2)$. The quadratic dependence of $\delta\dot{M}$ on
perturbation implies that the frequency of modulation of
${\dot M}$ is twice the frequency of the disk oscillation (see the
right panel of Fig.~\ref{fig:profiles}).


\section{The Inner Edge} \label{section-edge}


This is a very controversial point of a {\it fundamental}
importance. In Bohdan's own words \citep{afs-pac-2003}, ``Theory
of accretion disks is several decades old. With time ever more
sophisticated and more diverse models of accretion onto black
holes have been introduced. {\it However, when it comes to
modeling disk spectra, conventional steady state, geometrically
thin-disk models are still used, adopting the classical ``no
torque'' inner boundary condition at the marginally stable orbit
at the ISCO.}\citep[e.g.][]{bla-et-al-2001}. Recently, the no torque
condition for geometrically thin disks has been challenged by
several authors \citep{kro-1999, gam-1999, ago-kro-2000}. I did
not agree with their claim and presented simple arguments why the
no torque inner boundary condition is natural if an accretion disk
is geometrically thin \citep{pac-2000}, but the referee could not
be convinced. Thanks to the electronic preprint server, the paper
\citet{pac-2000} is readily accessible to all interested readers,
who can judge its validity.''

A more detailed quantitative analysis of the inner boundary
condition, refining arguments given by \citet{pac-2000} and making
them more precise, was done by \citet{afs-pac-2003}. They found
that \citet{kro-1999}, \citet{gam-1999}, and \citet{ago-kro-2000}
were qualitatively correct: a classical thin steady-state disk
must have some torque at the ISCO, but the effect is not as strong
as claimed by Krolik, Gammie and Agol.

Bohdan's line of argument is as follows:

\begin{itemize}

\item Today, the only models that make solid quantitative predictions
about the disk spectra that could be compared with observations
are the conventional {\it steady state}, geometrically thin-disk
models based on the classical ``no-torque'' inner boundary
condition at the marginally stable orbit at the ISCO.

\item It is not known today whether steady-state thin-disk type
accretion is possible at all. For serious technical reasons, not
even the most sophisticated 3D MHD numerical models of accretion
can describe very thin flows, or include radiative cooling.

\item Therefore, all arguments today concerning the behavior of the
steady-state flows near the ISCO are incomplete. They {\it all}
make simplifying assumptions.

\item The most fundamental property of the steady-state flows near
the ISCO is that they must necessarily be transonic (with the speed of
sound properly defined to include magnetic effects). The sonic
point is the {\it critical point} in the mathematical sense. It is
known from the theory of differential equations that certain
regularity conditions must by obeyed at a critical point.

\item Krolik, Agol and Gammie do not properly consider the
regularity conditions at the critical point(s). This is exactly
the reason why the stress they calculate at the ISCO is so large.
In all models which take a proper care of this important piece of
mathematics and calculate solutions which pass the critical points
smoothly, the stress at the ISCO is non-zero, but small.

\end{itemize}

A very important {\it numerical} work that supports
Paczy{\'n}ski's point of view was completed recently by
\citet{sha-2008}. It describes results of three-dimensional
general relativistic magnetohydrodynamical simulations of
geometrically thin accretion disk around a non-spinning black
hole. The disk has a relative thickness $h/r \sim 0.005-0.1$ over
the radius range $(2-20)GM/c^2$. In steady state, the specific
angular momentum profile of the inflowing magnetized gas deviates
by less than 2\% from that of the standard thin disk model with
the zero-inner-torque assumed. In addition, the magnetic torque at
ISCO is only $\sim$2\% of the inward flux of the angular momentum
at that radius. Both results indicate that magnetic coupling
across the inner edge in relatively unimportant for geometrically
thin disks around non-spinning black holes, which is in accordance
with Paczy{\'n}ski's ideas. However, until a mathematically
correct {\it analytic} model describing thin MHD accretion flows
near the ISCO becomes available, the controversy is likely to
continue.

\acknowledgements

This work was supported mostly by the Polish Ministry of Science
grant N203 009 31/1466, and by two smaller travel grants from
Princeton and Harvard universities.

\end{document}